\documentclass[preprint,sort&compress,12pt]{elsarticle}

\usepackage{bm} 
\usepackage{amsfonts} 

\newcommand\abm{{\ensuremath{\bm{a}}}}

\usepackage{tikz}
\usepackage{pgfplots}
\usepackage{pgfplotstable, booktabs}
\pgfplotsset{compat=1.9}

\usetikzlibrary{pgfplots.groupplots}
\usepgfplotslibrary{fillbetween}
\usetikzlibrary{calc,fit,matrix,arrows,automata,positioning,shapes}
\usetikzlibrary{arrows.meta}

\pgfplotsset{select coords between index/.style 2 args={
    x filter/.code={
        \ifnum\coordindex<#1\fi
        \ifnum\coordindex>#2\fi
    }
}}

\tikzset{
 invisible/.style={opacity=0},
 visible on/.style={alt={#1{}{invisible}}},
 alt/.code args={<#1>#2#3}{%
   \alt<#1>{\pgfkeysalso{#2}}{\pgfkeysalso{#3}}
 },
}

\usepackage{amssymb}
\usepackage{amsthm}
\usepackage{amsmath}
\usepackage{mathtools}
\usepackage{mathrsfs}
\usepackage[ruled, lined, linesnumbered, longend]{algorithm2e}
\SetKw{init}{Initialization:}
\let\oldnl\nl
\newcommand{\nonl}{\renewcommand{\nl}{\let\nl\oldnl}}
\usepackage{array}
\usepackage{multirow}
\usepackage{listings}
\usepackage{tabu}
\usepackage{enumerate}
\usepackage{lineno}
\usepackage{fullpage}
\usepackage{xcolor}
\usepackage[colorinlistoftodos]{todonotes}
\usepackage{bbm}
\usepackage[colorlinks=true]{hyperref}
\usepackage{url}
\usepackage{textcomp}
\usepackage{gensymb}
\usepackage{soul}
\usepackage{graphicx}
\usepackage{subfig}

\usepackage{fourier} 
\usepackage{array}
\usepackage{makecell}
\usepackage[export]{adjustbox}
\usepackage{float}
\usepackage{caption}
\usetikzlibrary{shapes}
\biboptions{numbers,comma,round,square}
\usepackage{arydshln}
\usepackage[para,online,flushleft]{threeparttable}
\usepackage{orcidlink}

\definecolor{myred}{RGB}{214,39,40}
\definecolor{mygray}{RGB}{176,176,176}
\definecolor{myorange}{RGB}{255,127,14}
\definecolor{myorangeT}{RGB}{221,171,100}
\definecolor{mygreen}{RGB}{44,160,44}
\definecolor{mylightgray}{RGB}{204,204,204}
\definecolor{mypurple}{RGB}{148,103,189}
\definecolor{mybrown}{RGB}{140,86,75}
\definecolor{steelblue}{RGB}{31,119,180}
\definecolor{intraray}{RGB}{127,193,219}
\definecolor{hemitrichs}{RGB}{26,74,93}

\DeclareRobustCommand\blueline{\raisebox{2pt}{\tikz{\draw[steelblue,line width = 0.8pt] (0,0)--(0.5,0);}}}
\DeclareRobustCommand\orangeline{\raisebox{2pt}{\tikz{\draw[myorange,line width = 0.8pt] (0,0)--(0.5,0);}}}

\DeclareRobustCommand\greenline{\raisebox{2pt}{\tikz{\draw[mygreen,line width = 0.8pt] (0,0)--(0.5,0);}}}
\DeclareRobustCommand\boldgreenline{\raisebox{2pt}{\tikz{\draw[mygreen,line width = 1.5pt] (0,0)--(0.6,0);}}}
\DeclareRobustCommand\redline{\raisebox{2pt}{\tikz{\draw[myred,line width = 0.8pt] (0,0)--(0.5,0);}}}

\DeclareRobustCommand\purpleline{\raisebox{2pt}{\tikz{\draw[mypurple,line width = 0.8pt] (0,0)--(0.5,0);}}}

\DeclareRobustCommand\blackdot{\raisebox{0.8pt}{\tikz{\draw[draw=none,fill=black] (0,0) circle (.4ex);}}}
\DeclareRobustCommand\orangedot{\raisebox{0.8pt}{\tikz{\draw[draw=none,fill=myorangeT] (0,0) circle (.4ex);}}}
\DeclareRobustCommand\bluedot{\raisebox{0.8pt}{\tikz{\draw[draw=none,fill=steelblue] (0,0) circle (.4ex);}}}

\theoremstyle{definition}

\theoremstyle{remark}

\newcommand{\hot}[1]{{\color{black} #1}}

\graphicspath{ {./figs/} }

\journal{Elsevier}
\DontPrintSemicolon
\begin{document}
\def\ps@pprintTitle{%
  \let\@oddhead\@empty
  \let\@evenhead\@empty
  \let\@oddfoot\@empty
  \let\@evenfoot\@oddfoot
}
\begin{frontmatter}

\title{Asynchronous Parallel Reinforcement Learning for Optimizing Propulsive Performance in Fin Ray Control}

\author[ndAME]{Xin-Yang Liu\orcidlink{0000-0003-1423-605X}\corref{contrib}}
\author[Maine]{Dariush Bodaghi\corref{contrib}}
\author[Maine,rit]{Qian Xue\orcidlink{0000-0002-8462-1271}}
\author[Maine,rit]{Xudong Zheng\orcidlink{0000-0001-5403-8862}\corref{corxh}}
\author[ndAME,ndLucy,ndEnergy]{Jian-Xun Wang\orcidlink{0000-0002-9030-1733}\corref{corxh}}%

\address[ndAME]{Department of Aerospace and Mechanical Engineering, University of Notre Dame, Notre Dame, IN}
\address[Maine]{
Department of Mechanical Engineering, University of Maine, ME}
\address[rit]{Department of Mechanical Engineering, Rochester Institute of Technology}
\address[ndLucy]{Lucy Family Institute for Data \& Society, University of Notre Dame, Notre Dame, IN}
\address[ndEnergy]{Center for Sustainable Energy (ND Energy), University of Notre Dame, Notre Dame, IN}
\cortext[contrib]{Authors contributed equally}
\cortext[corxh]{Corresponding author. Tel: +1 574-631-5302}
\ead{jwang33@nd.edu}

\begin{abstract}

Fish fin rays constitute a sophisticated control system for ray-finned fish, facilitating versatile locomotion within complex fluid environments. Despite extensive research on the kinematics and hydrodynamics of fish locomotion, the intricate control strategies in fin-ray actuation remain largely unexplored. While deep reinforcement learning (DRL) has demonstrated potential in managing complex nonlinear dynamics; its trial-and-error nature limits its application to problems involving computationally demanding environmental interactions. This study introduces a cutting-edge off-policy DRL algorithm, interacting with a fluid-structure interaction (FSI) environment to acquire intricate  fin-ray control strategies tailored for various propulsive performance objectives. To enhance training efficiency and enable scalable parallelism, an innovative asynchronous parallel training (APT) strategy is proposed, which fully decouples FSI environment interactions and policy/value network optimization. The results demonstrated the success of the proposed method in discovering optimal complex policies for fin- ray actuation control, resulting in a superior propulsive performance compared to the optimal sinusoidal actuation function identified through a parametric grid search. The merit and effectiveness of the APT approach are also showcased through comprehensive comparison with conventional DRL training strategies in  numerical experiments of controlling nonlinear dynamics. 
\end{abstract}

\end{frontmatter}

\section{Introduction}
Finned fish demonstrate extraordinary mobility by exploiting the innate flexibility and curvature of their body and fins, in contrast to the majority of man-made watercraft, which rely on propeller-driven propulsion. Through millions of years of evolutionary refinement, finned fish have developed oscillatory locomotion characterized by remarkable propulsion efficiency, maneuverability, and minimal noise generation~\cite{sfakiotakis1999review, sun2022recent}.
Despite extensive research on the kinematics and hydrodynamics of fish swimming over the years~\cite{sfakiotakis1999review, triantafyllou2000hydrodynamics, hermes2021bioinspired, cano2020key, videler1991fish}, the optimal control strategies for fin ray actuation largely remains elusive, primarily due to the intricate complexities arising from inherent flexibility and curvature of fish bodies and fins, coupled with their nonlinear interactions with the complex fluid environment.
Understanding these strategies is crucial for the development of bio-inspired soft robotic systems. While advances in computational fluid dynamics (CFD) and hydrodynamic experiments have enabled more detailed investigation into the underlying fluid-structure interaction (FSI) physics~\cite{lewin2003modelling, guglielmini2004propulsive, blondeaux2005numerical}, several challenges persist in comprehending the active control mechanism including (i) the highly nonlinear characteristics of the FSI system make the classic linearization-based control methods unsuitable, (ii) the continuum spatiotemporal and actuation parameter spaces result in an extremely high-dimensional control space, (iii) the considerable computational expense associated with simulating the interacting physics between flexible structures and complex fluid dynamics.

Deep reinforcement learning (DRL) has emerged as a promising approach for tackling highly non-linear dynamic control problems characterized by high-dimensional state-action spaces, as evidenced by recent advances in the field~\cite{mnih2015human, silver2016mastering, schrittwieser2020mastering, akkaya2019solving, badia2020agent57}. DRL leverages deep neural networks (DNNs) as the foundation for the control policy, enabling the agent to learn optimal actions through repeated interactions with the environment. In recent years, DRL has proven effective in managing various complex fluid dynamic systems across diverse scenarios, such as laminar and turbulent flows~\cite{rabault2019artificial, ghraieb2021single, ren2021applying, fan2020reinforcement, bucci2019control, beintema2020controlling, garnier2021review}, vortex shedding~\cite{rabault2019artificial, fan2020reinforcement, tang2020robust, rabault2020deep, ren2021applying, paris2021robust}, and fish swimming~\cite{gustavsson2017finding, verma2018efficient, zhu2021numerical, nair2022bio, novati2017synchronisation, gazzola2016learning, yan2020numerical,li2020vortex,zhu2022learning}. DRL's proficiency in handling high-dimensional control space can be attributed to its ability to learn complex mappings between states and actions through the use of DNNs. Furthermore, modern DRL techniques, such as experience replay and target networks~\cite{schaul2015prioritized,gao2017adaptive}, enhance stability and convergence during training, thereby improving its efficacy in addressing challenging control problems.
Despite the potential and initial successes of DRL in managing high-dimensional, non-linear systems, substantial challenges remain due to the high computational costs associated with high-fidelity (HF) simulations, particularly in the context of fluid-structure interactions. The trial-and-error nature of DRL requires a significant number of interactions with the environment, and each interaction involves numerically simulating FSI dynamics, such as in fish fin-ray control, making direct training of a DRL agent prohibitively expensive. Therefore, developing an efficient DRL solution capable of handling these computational demands is crucial for advancing the application of DRL to complex fluid/FSI dynamics.

To reduce the training cost of DRL in controlling fish locomotion or schooling, several studies have explored the utilization of fast surrogate models. This approach allows DRL agents to interact with these approximations, circumventing the necessity for direct training in computationally expensive HF simulated environments. A common practice involves leveraging low-fidelity (LF) numerical simulations, which rely on reduced dimensions and (over)simplified physics, providing a computationally efficient alternative for DRL training. For example, Gazzola et al.~\cite{gazzola2016learning} employed a pair of vortex dipoles to model swimmers, while Novati et al.~\cite{novati2017synchronisation} utilized a sinusoidal function to describe the swimmer's body curvature and prescribe the motion, avoiding the need for two-way coupled FSI. Some other studies directly neglected the shape of the swimmers and their influence on the surrounding fluids~\cite{colabrese2017flow, gustavsson2017finding}. Another promising strategy is to actively construct a DNN-based surrogate model for the environment during DRL training, known as model-based reinforcement learning (MBRL). The MBRL approach takes advantage of the fast inference speed of DNN surrogates, allowing numerous interactions with the learned environment. Notably, Liu et al.~\cite{liu2021physics} developed a physics-informed MBRL, introducing physics constraints in the MBRL training, leading to enhanced learning performance.

In LF simulated or DNN-learned environments, the numerous iterations required by DRL become manageable, and the learned policy will be subsequently applied to the target HF environment. For example, Verma et al.~\cite{verma2018efficient} utilized a two-dimensional (2D) LF numerical model to train the DRL agent, subsequently applying the learned policy to control a fish-like swimmer in the target HF environment based on three-dimensional (3D) direct numerical simulation (DNS). However, due to the notable differences between the training and target environments, the control policy obtained from the LF environment often falls short of achieving optimal performance in the target environment. While refining the LF-trained DRL agent in the target environment can enhance performance, the overall reduction in training costs, considering both the overhead of LF-based pre-training and subsequent fine-tuning in the target environment, remains a subject of debate. Some other studies have chosen to directly train their DRL agents using real experimental data~\cite{li2020vortex, fan2020reinforcement}, but this approach proves challenging for studying fish locomotion and swimming, given the impracticality or high difficulty associated with experimentally controlling real fish or manufacturing fish-like soft-body robots. Although previous work~\cite{shen2023bayesian}  illustrates the potential of utilizing DRL for experiment design, the challenges associated with training DRL models in real-world experiments persist. Therefore, direct training of a DRL agent in a computationally expensive HF simulated environment is sometimes necessary, particularly for studying fish fin-ray control involving complex nonlinear FSI dynamics, which are highly sensitive to the actions of the control agent.

To accelerate RL training in computationally demanding environments, a viable and effective approach is to simulate multiple environments concurrently. The success of this strategy has been demonstrated by Rabault et al.~\cite{rabault2019accelerating}. While the overall training time was reduced, indiscriminately running hundreds of environments can be inefficient, especially considering the heterogeneous hardware commonly employed in RL training. Furthermore, Rabault et al.~\cite{rabault2019accelerating} coupled parallel training environments with an on-policy DRL algorithm, Proximal Policy Optimization (PPO)~\cite{schulman2017proximal}. This method restricts training to interactions based on the current policy network, and its implementation requires the DRL agent to be trained after all the environments have completed their tasks. These constraints limit the potential advantages of running multiple environments in parallel. 
Previous studies have endeavored to enhance the suboptimal efficiency arising from concurrently simulating multiple environments, attributed to significant variance in simulation times across different environments by starting new agent-environment interactions in an asynchronous manner \cite{gu2017deep, mnih2016asynchronous}. However, it is important to note that despite simulating multiple environments asynchronously, the training of the policy network remains coupled with environment simulations. Consequently, the update of policy networks update is contingent on the completion of environment simulations, which leads to suboptimal efficiency in DRL training.

In this work, we propose a novel DRL training strategy, \textbf{A}synchronous \textbf{P}arallel \textbf{T}raining (APT), designed specifically to accelerate off-policy deep reinforcement learning efficiently and stably in computationally demanding environments, such as FSI dynamics for flexible fin-ray propulsion. APT revolves around the core concept of optimizing the utilization of heterogeneous hardware resources by harnessing asynchronous operations between CPUs and GPUs. By eliminating the need for synchronization between these computing units, APT effectively minimizes idle time and alleviates bottlenecks associated with conventional synchronous training approaches. This approach, in turn, significantly enhances overall training efficiency and speed in complex simulation environments, enabling faster convergence of the learning process. We successfully apply the APT method to two fin-ray control tasks: maximizing thrust and maximizing propulsion efficiency, achieving better performance compared to baseline methods. Our results illustrate the potential of APT as an effective solution for complex DRL tasks in computationally demanding scenarios. Additionally, we introduce a transfer learning-inspired technique named Global Searching and Local Fine-tuning (GSLF), designed to improve the performance and stability of DRL agents, particularly in the task of maximizing efficiency. The remainder of this paper is structured as follows. In Section~\ref{sec:method}, we provide a detailed description of the APT-based off-policy DRL methodology, outlining its key components and operation. Section~\ref{sec:results} presents the numerical results obtained for both the thrust maximization and efficiency maximization tasks. Further experimental findings on the performance of the APT method, along with an exploration of reward function choices, are discussed in Section~\ref{sec:dis}. Finally, Section~\ref{sec:conclusion} concludes the paper.



\section{Methodology}

\label{sec:method}

\subsection{The simulated FSI environment}

\begin{figure}[!htp]
\centering
\hfill
\subfloat[]{\includegraphics[width=0.35\textwidth]{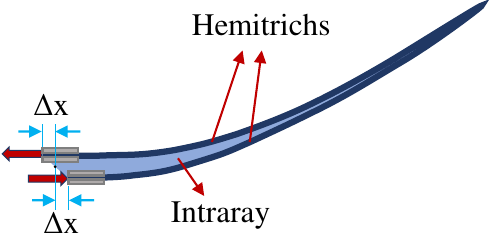}}
\hfill
\subfloat[]{\includegraphics[width=0.4\textwidth]{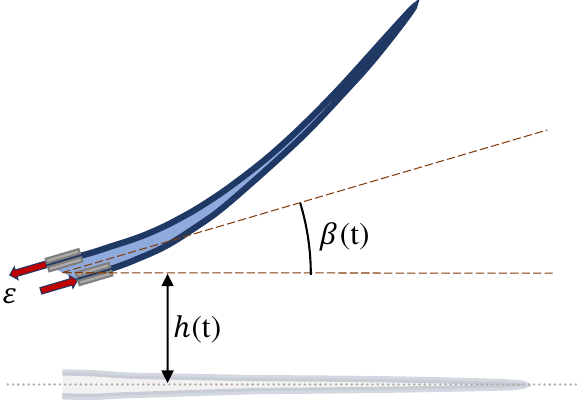}}
\hfill.
\caption{Schematics of (a) the fin-ray deformation with muscle actuation by applying offset of $\varepsilon$; (b) the fin-ray root motions of pitching, plunging, and muscle actuation}
\label{fig:env_discribe}
\end{figure}

In this work, we employ DRL to explore control strategies for fin-ray actuation in the fish-fin propulsion within a simulated FSI environment. Illustrated in Fig.~\ref{fig:env_discribe}(a), fish fin ray is characterized by a unique bilaminar structure, consisting of the intraray, made primarily of soft tissue, and the bony hemitrichs that encapsulate the intraray. The bilaminar nature of this structure enables real-time control of each ray through antagonistic muscle actuation at the base of the ray, causing a displacement offset of two hemitrichs. This mechanism allows for the generation of intricate stiffness and curvature variations across the entire fin in space and time.
In the simulated FSI environment, the muscle actuation is represented by applying the offset of $\varepsilon = \Delta x/L$ to the root of each hemitrich, where $\Delta x$ is the root displacement of the hemitrich and $L = 4 cm$ is the length of the fin ray. The detailed material properties and dimensions of the ray model can be found in~\cite{bodaghi2023effects}.

To faithfully replicate the biomechanical dynamics observed in natural fish locomotion, our simulation incorporates prescribed pitching and plunging motions within the fin ray model. As depicted in Fig.~\ref{fig:env_discribe}(b), the kinematics of the ray are governed by a synergistic interaction between the root's pitching-plunging motions and subsequent bending introduced by the root displacement $\varepsilon$. This kinematic scheme is informed by high-resolution photogrammetric analyses of fish swimming~\cite{liu2019image}, which have revealed that the fin-ray root undergoes periodic motions described by the functions $\beta(t)$ for pitching and $h(t)$ for plunging, occurring with a 90-degree phase shift, 
\begin{equation}
\begin{split}
\beta(t) &= \beta_0 \sin{(2\pi f t)},\\
h(t)&= h_0 \sin{(2\pi f t + \pi/2)},
\end{split}
\label{eq:prescribedMotion}
\end{equation}
where two constants, $\beta_0 = 0.392$ and $h_0 = 0.25$ are derived from the photogrammetry~\cite{liu2019image}; the beating frequency is denoted as $f=2$ Hz. The upstream flow velocity is set as $v=10 \mathrm{{cm\cdot s^{-1}}}$, resulting in a Strouhal number ($St = 2 h_0 f/v$) of 0.4, falling within the natural range ($0.2 < St <0.4$) typically observed in aquatic environment~\cite{anderson1998oscillating}. Additionally, a kinematic viscosity value of $\nu = 1.084\times10^{-6}$ is chosen, yielding a Reynolds number ($Re = vL/\nu$) of 3690, which is in moderate range of fish swimming with a high viscous effect~\cite{bone2008biology}.

In the current study, the simulation environment employs an in-house FSI solver which couples a sharp interface immersed boundary method based incompressible flow solver with a finite element method based solid dynamics solver~\cite{zheng2010coupled}. This flow solver incorporates a multi-dimensional ghost-cell methodology, adept at handling the complexities of moving boundaries with a second-order accuracy both globally and in proximity to the immersed boundary.  
\begin{figure}[!htp]
\centering
\includegraphics[width=0.8\textwidth]{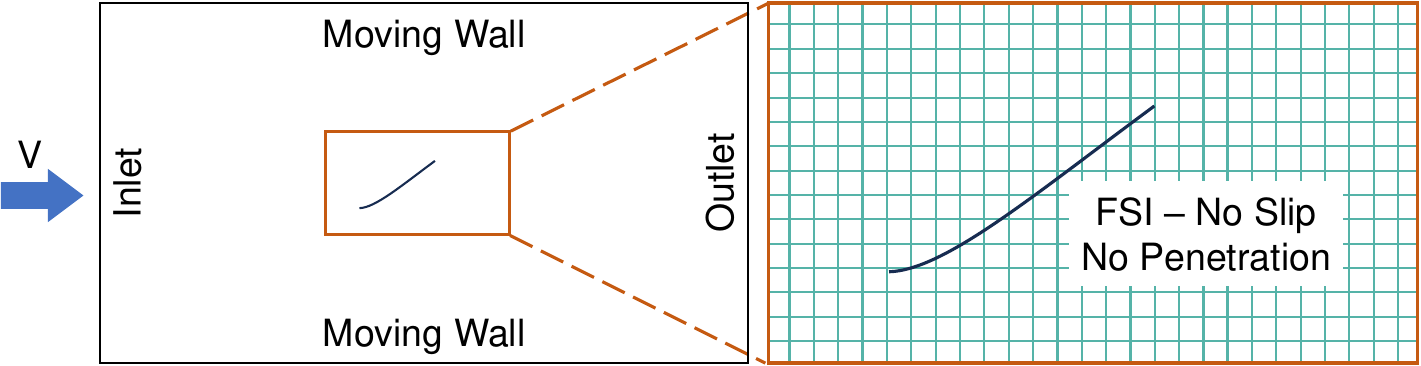}
\caption{Boundary conditions of the flow solver and near body computational grids.}
\label{fig:numerical}
\end{figure}
Figure~\ref{fig:numerical} illustrates the computational domain and boundary conditions of the simulated FSI environment. The domain is discretized using a grid of $112 \times 97$ Cartesian cells with the finest grid size of $0.038L \times 0.038L$ near the fin-ray to resolve the near-field vortex structures.  The left boundary is set as a velocity inlet with an upstream velocity $v$, while the top and bottom boundaries are treated as moving walls, synchronized with the velocity $v$. The right boundary is defined with a zero pressure and zero velocity gradient condition. For detailed simulation setup and FSI solver, please refer to~\cite{bodaghi2023effects}.

\subsection{Deep reinforcement learning}
In reinforcement learning (RL), a RL agent is tasked with learning an optimal control strategy or policy $\pi$ from its experience of interacting with the environment. This learning process involves the agent iteratively interacting with its environment and making decisions ($\bm{a}_i$) at each control step ($i$) based on its observations ($\bm{o}_i$) of the current environment state ($\bm{s}_i$). After executing an control action ($\bm{a}_i$), the environment returns the new state ($\bm{s}_{i+1}$) and a reward ($r_i$), which serves as a feedback signal for the action taken. This interaction process can be mathematically described as follows,
\begin{align}
    \bm{a}_i &= \pi(\bm{o}_i), \\
    \bm{o}_i &= f_O(\bm{s}_i), \\
    \bm{s}_{i+1} &= \bm{\mathcal{F}}(\bm{s}_i;\, \bm{a}_i), \label{eq:env}\\
    r_{i} &= f_r(\bm{s}_{i+1};\, \bm{a}_i),
\end{align}
where $f_O$ denotes the observation function, $f_r$ is the reward function, and $\mathcal{F}$ represents the dynamics of the environment. The objective of the RL agent is to maximize the cumulative reward over an episode, which is typically composed of a sequence of control steps. 

In the context of deep reinforcement learning (DRL), the policy $\pi$ is learned by deep neural networks, formulated as,
\begin{equation}
    \bm{a}_i = \pi_{\bm{\theta}}(\bm{o}_i) = \pi(f_O(\bm{s}_i); \,\bm{\theta})
    \label{eq:policy}
\end{equation}
where $\pi_{\bm{\theta}}$ symbolizes the policy network parameterized by trainable weights $\bm{\theta}$. The training in DRL is an optimization problem aimed at maximizing the expected cumulative reward, expressed as,
\begin{equation}
    \max_{\bm{\theta}}\,R = \max_{\bm{\theta}}\sum_{i=1}^{N} \gamma^{i-1} r_i = \max_{\bm{\theta}}\sum_{i=1}^{N} \gamma^{i-1} f_r(\bm{s}_{i+1};\,\bm{a}_i) = \max_{\bm{\theta}}\sum_{i=1}^{N} \gamma^{i-1} f_r\Big(\bm{s}_{i+1};\,\pi_{\bm{\theta}}\big(f_O(\bm{s}_i)\big)\Big),
    \label{eq:rlreturn}
\end{equation}
where $R$ is the expected return, and $\gamma$, the discount factor between 0 and 1, reflects the preference for immediate rewards over future rewards. In our implementation, we adopt $\gamma=0.99$, aligning with standard practices in DRL~\cite{haarnoja2018soft,christodoulou2019soft}. In general, DRL algorithms can be divided into two categories: on-policy and off-policy, based on the source of interaction data used for updating the neural networks. On-policy algorithms rely on data derived from the current policy, whereas off-policy algorithms utilize historical experiences, which typically results in greater sample efficiency. The focus of the proposed method is to enhance the training efficiency of off-policy algorithms further. Off-policy algorithms are characterized by their use of a replay buffer, denoted as $\mathcal{D}$, to store past interaction experiences $e_i$. Each interaction experience in this context is a tuple comprising the current state $\bm{s}_i$, the action $\bm{a}_i$ taken by the RL agent based on this state, the next-step state $\bm{s}_{i+1}$ resulting from the action, and the associated reward $r_i$, formally represented as $e_i = (\bm{s}_i, \bm{a}_i, \bm{s}_{i+1},r_i)$. The neural networks are then updated using the data accumulated in the replay buffer, as detailed in Algorithm~\ref{alg:conventional}.
\begin{algorithm}[!t]
\caption{Conventional off-policy reinforcement learning}\label{alg:conventional}
\SetAlgoLined
\nonl\init{\textup{Initialize the neural network(s) with trainable parameters $\theta$; empty the replay buffer $\mathcal{D}$; initialize the environment state $\bm{s}_0$}}\\
\While{not converge}{
\For{$i =1:n_i$}{
Get the observation $\bm{o}_i$ of the environment state $\bm{s}_i$\\
Interact with the environment based on the current learned strategy $\bm{a}_i = \pi(\bm{s}_i)$\\
Get the observation $\bm{o}_{i+1}$ of the new environment state $\bm{s}_{i+1}$; get the reward $r_i$\\
Store the interaction $e_i = (\bm{o}_i, \bm{a}_i, \bm{o}_{i+1},r_i)$ to replay buffer $\mathcal{D}$\\
\If{episode ends}{reset the environment $\bm{s}_0$}
}
Sample a batch of experiences $\mathcal{B} = \{e_i|e_i \in \mathcal{D}\}$ \\
Update the neural network(s) parameters $\theta$ based on the data $\mathcal{B}$}
\end{algorithm}

\subsection{Enhancing RL Training Efficiency through Asynchronous Parallel Training}
\begin{figure}[!t]
    \centering
    \subfloat[Conventional DRL Training]{\includegraphics[width=\textwidth]{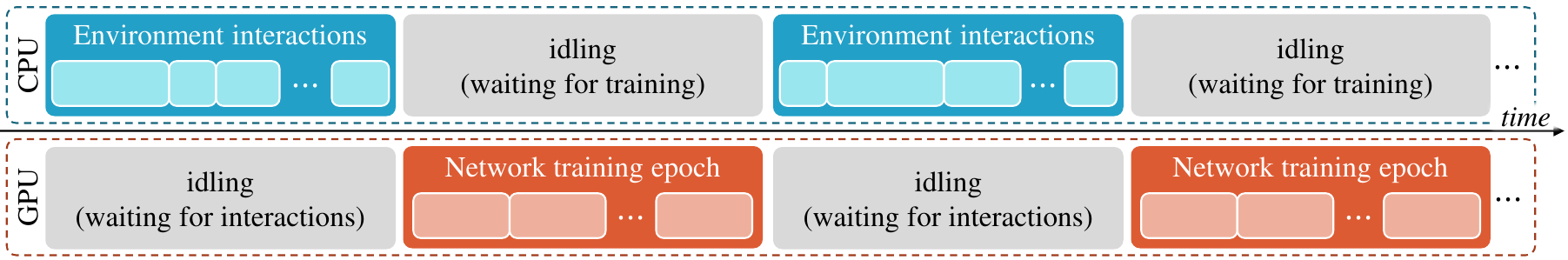}\label{fig:sync_s:stru}}\\
    \subfloat[Conventional Parallel DRL Training]{\includegraphics[width=\textwidth]{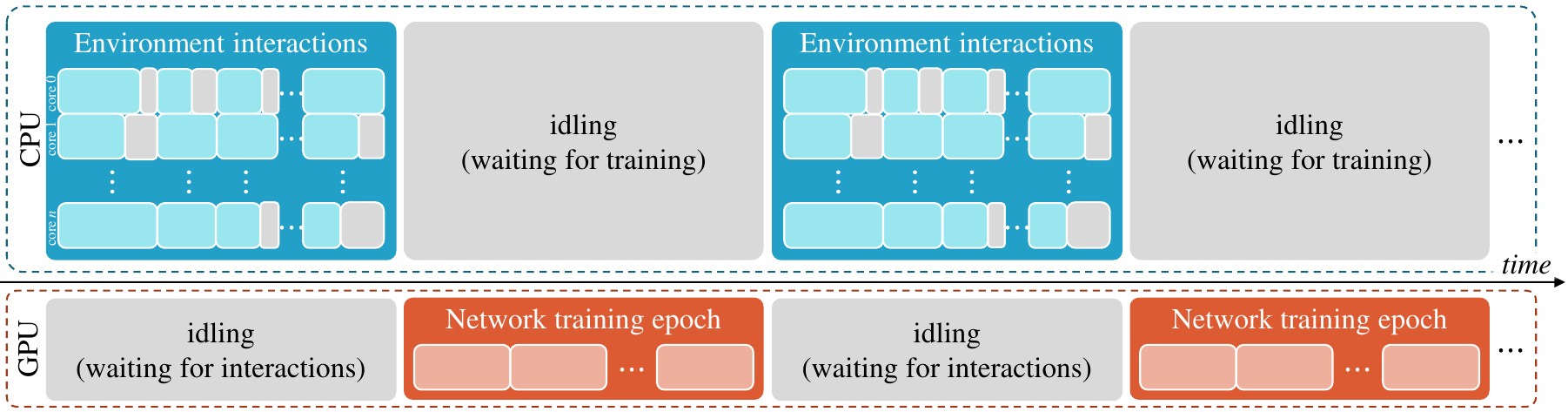}}\\
    \subfloat[Asynchronous Parallel Training (APT)]{\includegraphics[width=\textwidth]{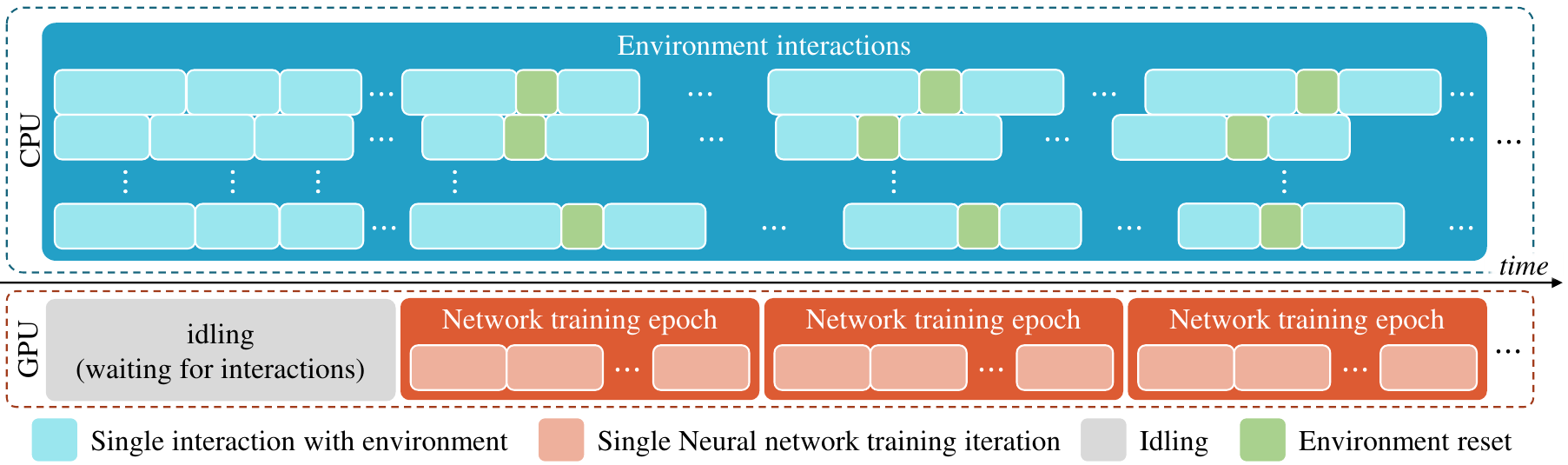}}
    \caption{Time consumption schematics of 3 different RL training strategies}
    \label{fig:comparison}
\end{figure}

In modern DRL frameworks, the training typically involves a division of labor between CPUs and GPUs. CPUs are generally tasked with simulating the environment dynamics, while GPUs are dedicated to the process of updating neural network parameters. Although these heterogeneous computing units are used, conventional DRL algorithms predominantly adhere to a synchronized operational model, wherein CPUs and GPUs alternate their activities, leading to periods of inactivity as one waits for the other to complete its task, as depicted in Fig.~\ref{fig:comparison}(a).
This synchronized approach results in suboptimal utilization of the heterogeneous hardware, introducing significant latency and inefficiencies, thereby limiting the overall performance and leading to the underutilization of computing resources during synchronization periods. Running multiple environments in parallel can enhance CPU utilization for certain tasks, as noted in~\cite{rabault2019accelerating}. However, this method still remain inefficient, particularly when the time cost associated with different environments varies significantly, a common scenario in simulating complex FSI problems, as shown in Fig.~\ref{fig:comparison}(b). This inefficiency persists even in scenarios where only CPUs are employed for both environment interaction and neural network training, as the processing time is dictated by the slowest environment.

To mitigate these inefficiencies and to leverage the full potential of heterogeneous computing systems in DRL, we introduce the Asynchronous Parallel Training (APT) algorithm. 
\begin{algorithm}[!htp]
\caption{Asynchronous Parallel Training (APT) for off-policy reinforcement learning}\label{alg:apt}
\SetAlgoLined
\SetKwBlock{Parallel}{do in parallel}{end}
\nonl\init{\textup{Initialize the neural network(s) with trainable parameters $\theta$; empty the replay buffer $\mathcal{D}$; initialize all the environment $\mathrm{env}_j$, with state $\bm{s}_0^j\,$ (where $\bm{s}_i^j$ is the state of $\mathrm{env}_j$, $j=0,1,2,3\cdots\,n$)}}\;
\While{not converge}{
    \Parallel{
    \textit{interact\_with\_env}\,(\,$\mathrm{env}_0$\,)\;
    \textit{interact\_with\_env}\,(\,$\mathrm{env}_1$\,)\;
    $\cdots$\;
    \textit{interact\_with\_env}\,(\,$\mathrm{env}_n$\,)\;
    
    Update neural network(s) trainable parameters $\theta$ based on data sampled from $\mathcal{D}$
    }}
    Send terminate signal to all \textit{interact\_with\_env} processes. \;
    
    \SetKwProg{Def}{def}{:}{}
    \Def{interact\_with\_env\,\textup{(\,$\mathrm{env}_j$)}}{
        \While{\textup{main process not terminated}}{
            Get the observation $\bm{o}_i^j$ of the state $\bm{s}_i^j$ of environment $\mathrm{env}_j$\;
            Interact with $\mathrm{env}_j$ based on the current learned strategy $\bm{a}^j_i = \pi(\bm{o}_i^j)$\;
            Ge the observation $\bm{o}_{i+1}^j$ of new environment state $\bm{s}_{i+1}^j$; get the reward $r_i^j$\;
            Store the interaction $e^j_i = (\bm{o}_i^j, \bm{a}_i^j, \bm{o}_{i+1}^j,r_i^j)$ to replay buffer $\mathcal{D}$\;
            \If{episode ends}{reset state in $\mathrm{env}_j$ $\bm{s}_0^j$}
        }
    }
\end{algorithm}
APT overhauls the training process by decoupling environment simulation, performed by CPUs, from neural network training, carried out by GPUs. This strategy is depicted in Fig~\ref{fig:comparison}(c), where the asynchronous nature of APT is evident: CPUs continuously simulate multiple environment interactions without waiting for GPUs to complete training epochs, and vice versa. This approach enables simultaneous operations, eliminating idle times that previously characterized CPU-GPU interdependence. The asynchronous operation allows for a non-blocking workflow where CPUs can process subsequent environment interactions while GPUs concurrently optimize neural network parameters, leading to a significant reduction in total training time and maximizing the utilization of available computational resources. APT ensures active engagement of CPUs and GPUs, enhancing the training pipeline's efficiency. The implementation details of APT, which include the scheduling of tasks between CPUs and GPUs, the management of the replay buffer, and the updating protocols for neural networks, are comprehensively detailed in Algorithm~\ref{alg:apt}. In the presented APT framework, environment resets are handled independently of the main training loop, allowing for uninterrupted environment simulations and network training sessions. This is particularly beneficial when dealing with complex FSI simulations, where the computational load can vary significantly. We demonstrate APT's efficacy using the state-of-the-art off-policy DRL algorithm, Soft Actor-Critic (SAC)~\cite{haarnoja2018soft}. Nonetheless, APT is not exclusive to any specific DRL algorithm; it is universally adaptable to various off-policy algorithms with experience replay mechanisms, such as Deep deterministic policy gradient (DDPG) ~\cite{lillicrap2015continuous} and Twin delayed DDPG (TD3)~\cite{fujimoto2018addressing}.

\section{Numerical Experiments and Results}
\label{sec:results}

\subsection{Problem formulation and DRL setting}

\begin{figure}[htp!]
    \centering
    \hspace*{\fill}
    \subfloat[$\bm{o}_{flow}$]{\includegraphics[width=0.52\textwidth,valign=c]{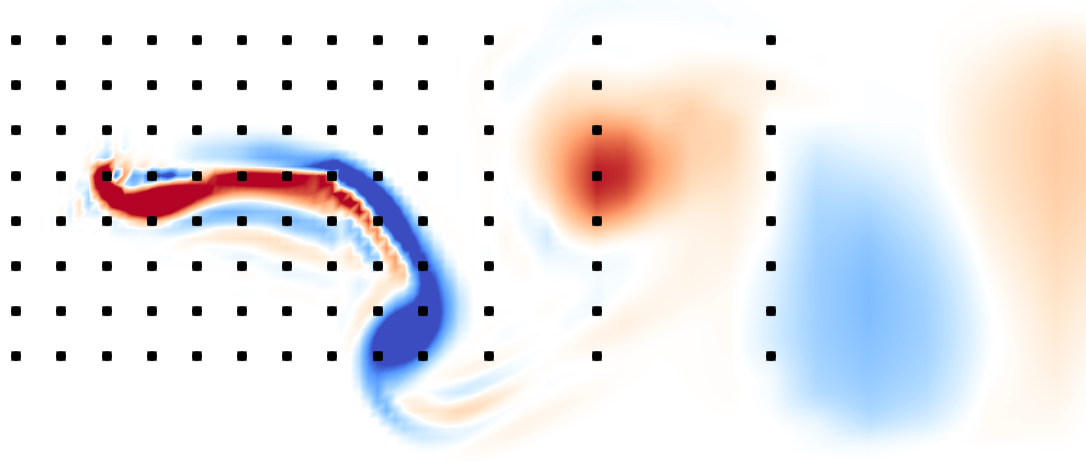}}
    \hfill
    \hfill
    \subfloat[$\bm{o}_{fin}$]{\includegraphics[width=0.28\textwidth,valign=c]{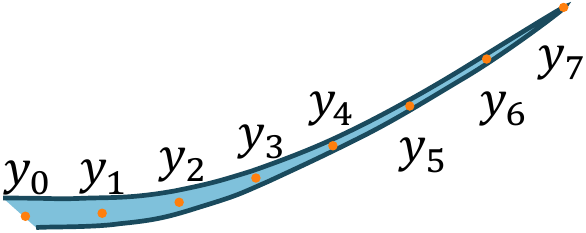} \vphantom{\includegraphics[width=0.52\textwidth,valign=c]{figs/probes.png}}}
    \hspace*{\fill}
    \caption{Illustration of the control Parameter and the observation space for the DRL agent. (a) depicts the observation vector of surrounding flow $\bm{o}_{flow}$, which includes the stream-wise velocity $u_x$ probed at the locations indicated by black dots (\blackdot); (b) visualizes the observation vector of fin-ray deformation $\bm{o}_{fin}$, comprising the $y$-coordinates of eight equidistant points (\orangedot) along the fish fin ray. }
    \label{fig:obs_space}
\end{figure}

\paragraph{Observation Space}
To approximate real-world conditions for a fish or fish-like robot, the RL agent's observation space is confined to the immediate flow field around the fin. Specifically, the observation space $\mathbb{O} \subset \mathbb{S}$ is a subset of the full state space $\mathbb{S}$. Namely, the observed state is composed of the $x$-direction velocity captured by an array of 104 probes, denoted as $\bm{o}_{flow} \in \mathbb{R}^{104}$, depicted in Fig.~\ref{fig:obs_space}(a). It also includes an 8-dimensional state vector $\bm{o}_{fin} \in \mathbb{R}^{8}$, describing the deformation status of the fin ray at each control step. The complete observation space is thus given by,
\begin{equation}
    \mathbb{O}: \{\bm{o}\} = \left\{\bm{o}_{flow};\, \bm{o}_{fin}\right\} \in \mathbb{R}^{112}
\end{equation}

\paragraph{Action space}
The RL agent modulates the root displacement $\varepsilon$ of the fin ray to achieve the control objective, as illustrated in Fig.~\ref{fig:env_discribe}). Due to realistic consideration, the action $a_i\in \mathbb{R}$ is subject to the following constraint, 
\begin{equation}
    a_i = \varepsilon_i - \varepsilon_{i-1} \in  (-3.14\times 10^{-5}n, 3.14\times 10^{-5}n)
\end{equation}
where $i$ indexes the current control step, and $n = 50$ is the number of numerical steps in one control step. The action $a_i$ is evenly distributed across each numerical step within the $i^\mathrm{th}$ control step,
\begin{equation}
    \alpha_{i} = a_i/n
\end{equation}
where $\alpha_{i}$ represents the incremental displacement at each numerical step. The chosen action $a_i$ is determined by the policy network $\pi$ with parameters $\theta$ and current observation $\bm{o}_i$.

\paragraph{Neural Network Architecture}
Our DRL model employs two key neural networks: the policy network, which determines the agent's actions, and the Q-function network, which estimates the value of action-state pairs. Both networks are constructed as multilayer perceptrons (MLPs) featuring two hidden layers. Each hidden layer is densely populated with 512 neurons, ensuring a robust capacity for learning complex representations of the environment and action spaces. For non-linear transformation within the hidden layers, the Rectified Linear Unit (ReLU) activation function is applied. The output layer of the Q-function network utilizes the identity activation function, providing a direct linear output that correlates with the expected returns of the state-action pairs. On the other hand, the policy network's output layer employs the hyperbolic tangent (\textit{Tanh}) activation function. The use of \textit{Tanh} is particularly crucial as it bounds the output, ensuring that the actions generated by the policy net are confined within the predefined valid range. These architectural choices for the neural networks are designed to balance computational efficiency with the ability to capture the complexity of the control task at hand, ultimately leading to more effective and realistic policy development within the constraints of the modeled environment. 

\paragraph{Episode and control step}
An episode with a duration $T = 8\mathrm{s}$ is segmented equally into $N = 80$ control steps. Each step lasts $\tau = T/N$ in time and consists of $n = 50$ numerical steps in order to keep the control frequency within a practical range. Within each episode, two prescribed motions are applied to the fin ray in addition to the root displacement $\varepsilon$: the translational movement $h(t)$ and rotation movement $\theta(t)$ as defined in Eq.\ref{eq:prescribedMotion}. One episode corresponds to four cycles of these prescribed motions.

\subsection{Baseline control method for comparative analysis}
To comprehensively evaluate our DRL strategy, we first formulated a baseline control method for comparative purpose. This baseline leverages prescribed motions $h(t)$ and $\beta(t)$, characterized by trigonometric functions. Intuitively, we propose that the optimal control strategy for the fin-ray displacement $\varepsilon(t)$ might adhere to a sinusoidal pattern,
\begin{equation}
\varepsilon(t) = \varepsilon_0 \sin{(2\pi f t + \varphi)},
\end{equation}
where $\varepsilon_0$ is the amplitude of displacement, and $\phi$ is the phase shift, both of which are pivotal parameters that are posited to significantly impact the propulsive effectiveness of the fin ray. By introducing a sinusoidal control strategy, the high-dimensional spatiotemporal control space can be substantially simplified into a two-parameter sinusoidal function space, enabling the application of traditional optimization techniques, including an exhaustive grid search, to systematically explore the parameter space and identify the parameters that yield optimal propulsion.

In pursuit of identifying the most effective parameter set, we conducted a systematic grid search within the two-dimensional parameter space. The amplitude $\varepsilon_0$ varied from 0.0002 to 0.007 in increments of 0.0004, and the phase shift $\phi$ was adjusted from 0° to 315° at 45° intervals. This approach resulted in 144 distinct scenarios. The performance of each scenario was evaluated based on critical metrics that reflect the propulsive efficiency and control effectiveness. The scenario exhibiting superior performance was selected as the benchmark for comparison. This carefully optimized sinusoidal control strategy provides a direct and relevant comparison for assessing the advantages brought forth by the DRL-controlled approach, thereby validating the improvements in control strategy derived from DRL optimization. More details about the parametric analysis of the fin ray actuation in functional space of sinusoidal movements can be found in our previous work~\cite{bodaghi2023effects}.

\subsection{Maximize thrust}
In the first case, the DRL agent is to learn a control policy $\pi^T_\theta$ that maximizes the accumulated thrust $F_T$ produced over a single episode. The optimization goal is formulated as,
\begin{equation}
\max_{a_i\sim\pi^T_{\theta}}F_T =\max_{a_i\sim\pi^T_{\theta}}\sum_{i=1}^N F_{T, i} = \max_{a_i\sim\pi^T_{\theta}}\sum_{i=1}^{N} \int_{t_{i-1}}^{t_i} f_T\Big(\bm{s}_i(t),a_i\Big) \mathrm{d}t,
\label{eq:maxT:goal}
\end{equation}
where $F_{T,i}$ represents the \hot{accumulated thrust} within each control step, while $f_T$ represents the instantaneous thrust. $N$ is the total number of control steps within one episode, and $t_i - t_{i-1} = \tau$ is the time duration of the $i^{\rm{th}}$ control step, where $\tau$ is a constant. Accordingly, the reward $r\left(\bm{s}_i, a_i\right)$ can be defined as the \hot{thrust} $F_{T,i}$ at control step $i$, 
\begin{equation}
    r\left(\bm{s}_i, a_i\right) = F_{T,i} = \int_{t_{i-1}}^{t_i} f_T\Big(\bm{s}_i(t),\,a_i\Big) \mathrm{d}t.
\end{equation}
Employing our APT methodology, the SAC algorithm guided the DRL agent to an optimal policy $\bm{\pi}^T$, aiming to maximize thrust. The agent reached this optimal policy after $6\times 10^4$ interactions with the environment, producing a total thrust of $3.2682\times10^4\mathrm{N\cdot s}$, nearly double that achieved by the baseline optimal control method, $1.7519\times10^4\mathrm{N\cdot s}$. 

\begin{figure}[!htp]
    \centering
    \subfloat{\includegraphics[width=0.48\textwidth,]{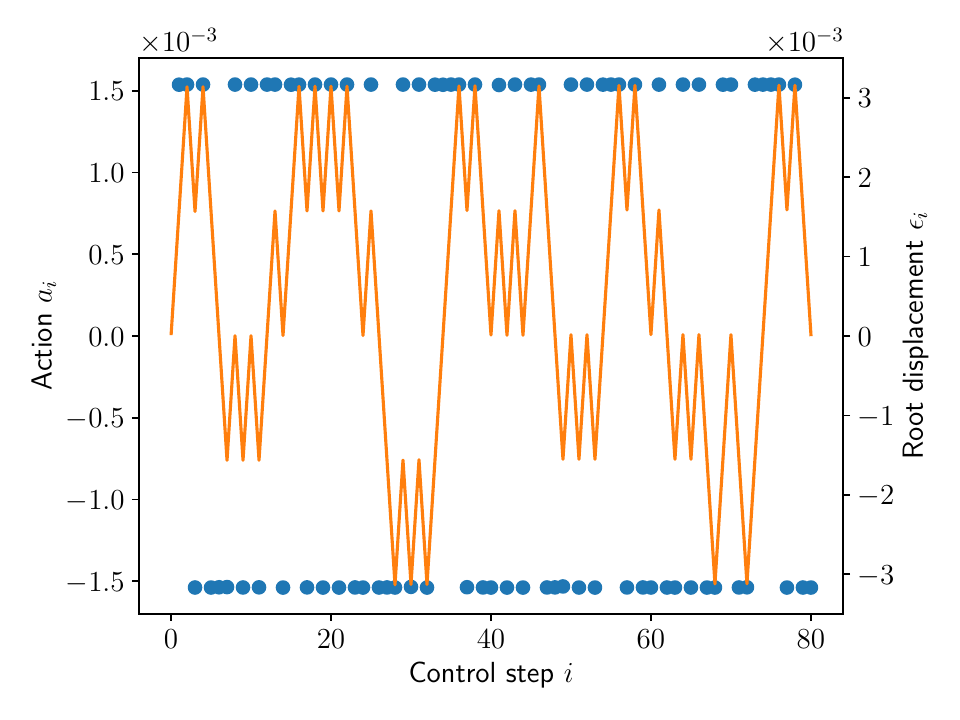}}
    \subfloat{\includegraphics[width=0.48\textwidth]{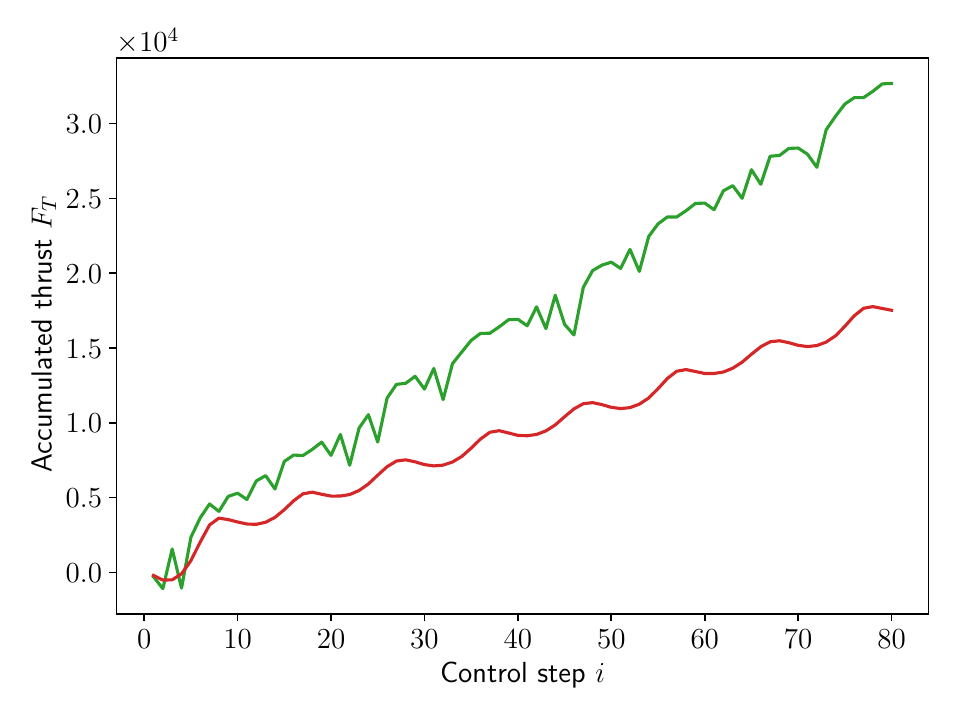}}
    \vspace{-12pt}
    \caption{Left panel: the actions $a_i$ (\bluedot) taken by the DRL agent, the corresponding root displacement $\varepsilon$ (\orangeline), and the accumulated thrust $F_T$ (\greenline) generated by the DRL-controlled fin ray. Right panel: the time series of thrust generated by the max-thrust DRL agent compared with that obtained by baseline method (\redline) during one episode.}
    \label{fig:MaxT:act}
\end{figure}

Figure~\ref{fig:MaxT:act} illustrates the dynamic control behavior of the DRL agent, captured through the actions taken $a_i$, the root displacement of the fin ray $\varepsilon_i$, and the resultant \hot{cumulative thrust} $F_T$ generated over the course of a single control episode. 
As observed in the left panel, the DRL agent consistently selects actions of maximum magnitude across all control steps, leading to a pronounced series of peaks and troughs in the root displacement ($\varepsilon_i$) profile, signaling a aggressive control policy tailored for optimized thrust generation. The right panel demonstrate the success of this strategy, as evidenced by the steadily climbing \hot{cumulative thrust} ($F_t$) curve, with only minor perturbations due to the periodic nature of the prescribed translational and rotational motions. This pattern indicates that the DRL agent has learned to effectively mitigate adverse factors and fully exploit the available action space, thereby maximizing thrust output throughout the episode. The consistent upward trajectory of the $F_t$ curve, particularly when contrasted with the thrust generated by the baseline method, validates the DRL agent's capability to dynamically adjust and improve its control policy, effectively boosting the total thrust.

\begin{figure}[!t]
    \centering
    
    \subfloat[The $60^{th}$ control step (DRL)]{\includegraphics[width=0.33\textwidth]{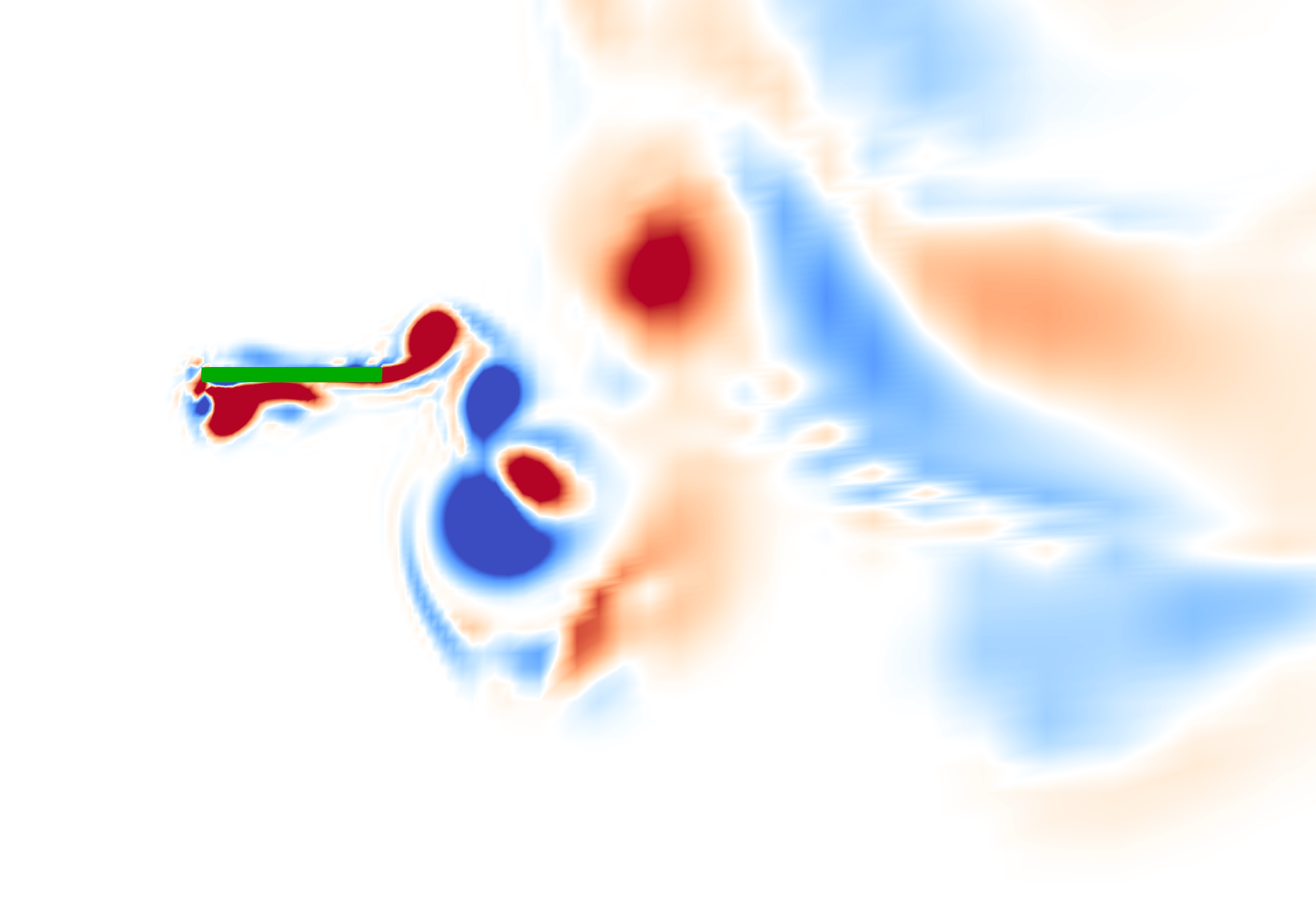}}
    \subfloat[The $65^{th}$ control step (DRL)]{\includegraphics[width=0.33\textwidth]{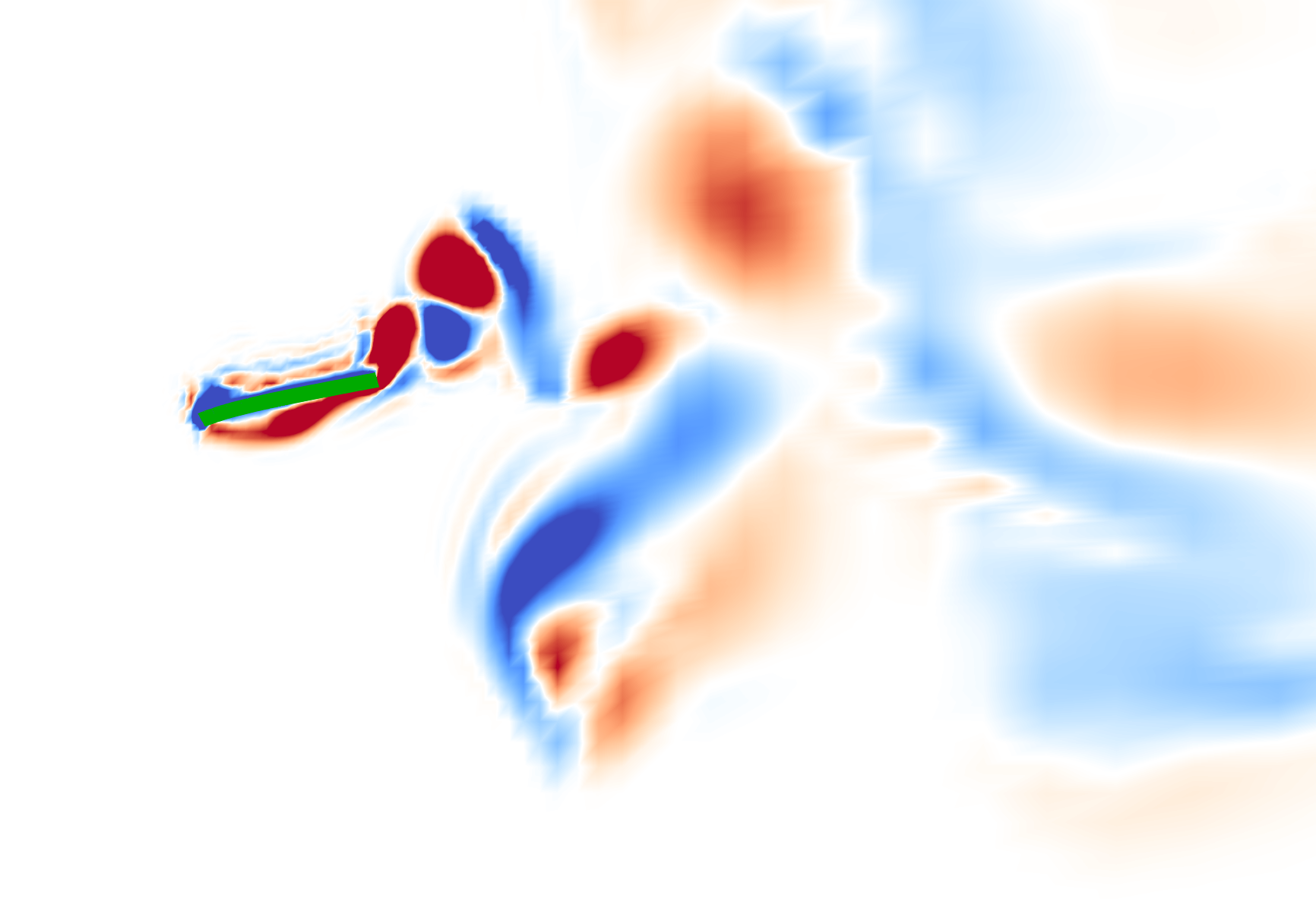}}
    \subfloat[The $70^{th}$ control step (DRL)]{\includegraphics[width=0.33\textwidth]{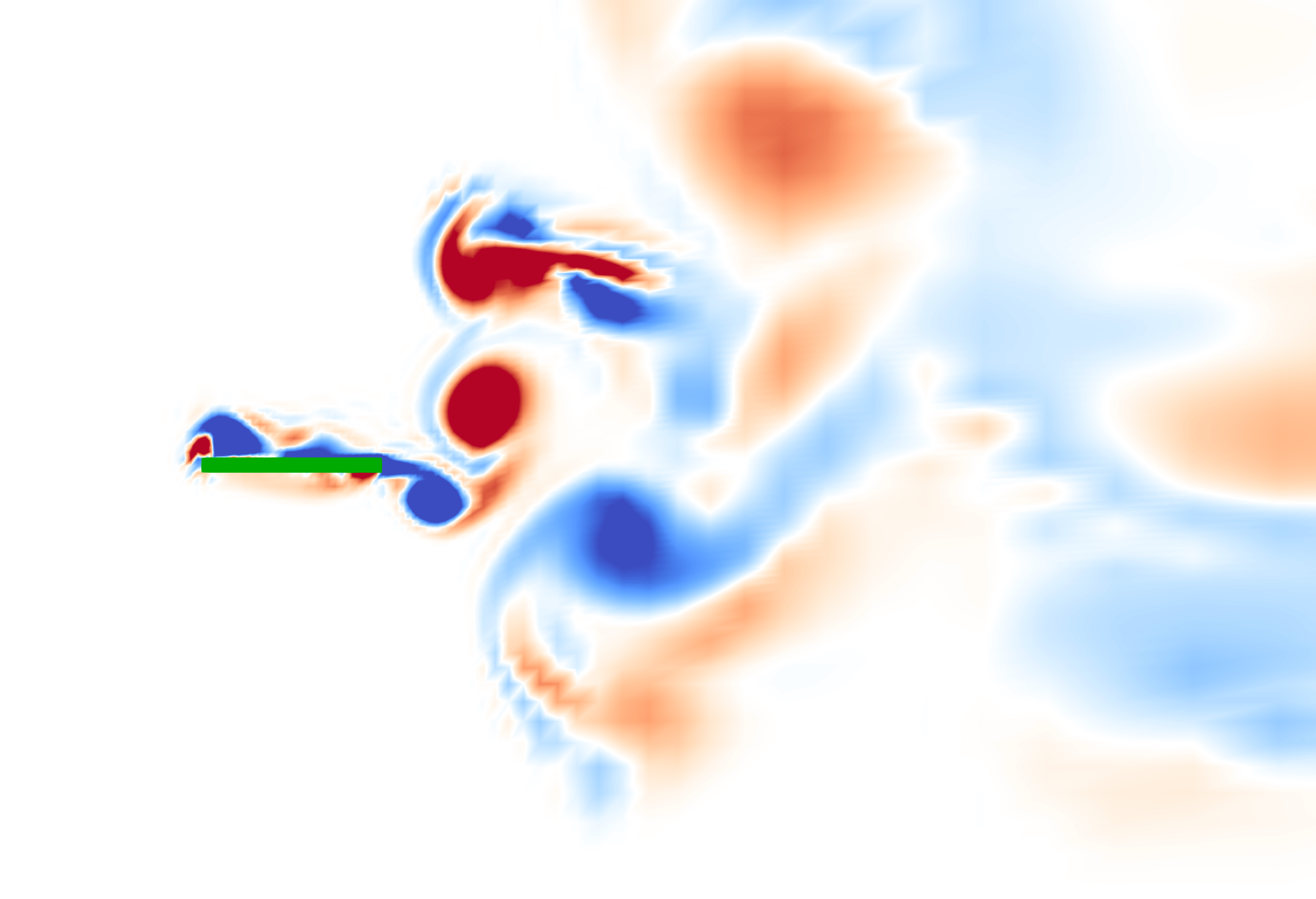}}
    
    \vspace{-2ex}

    \subfloat[The $75^{th}$ control step (DRL)]{\includegraphics[width=0.33\textwidth]{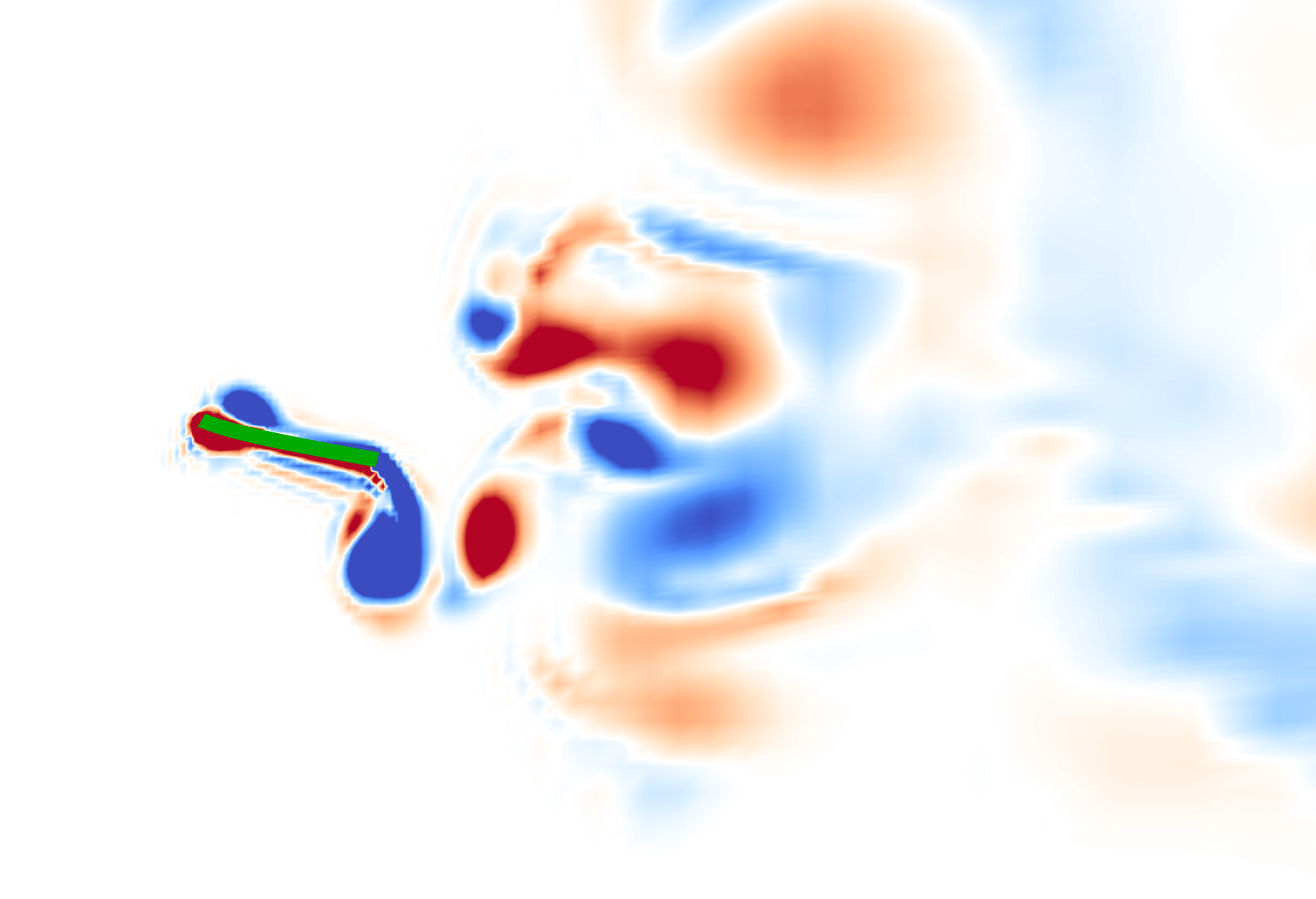}}
    \subfloat[Last control step (DRL)]{\includegraphics[width=0.33\textwidth]{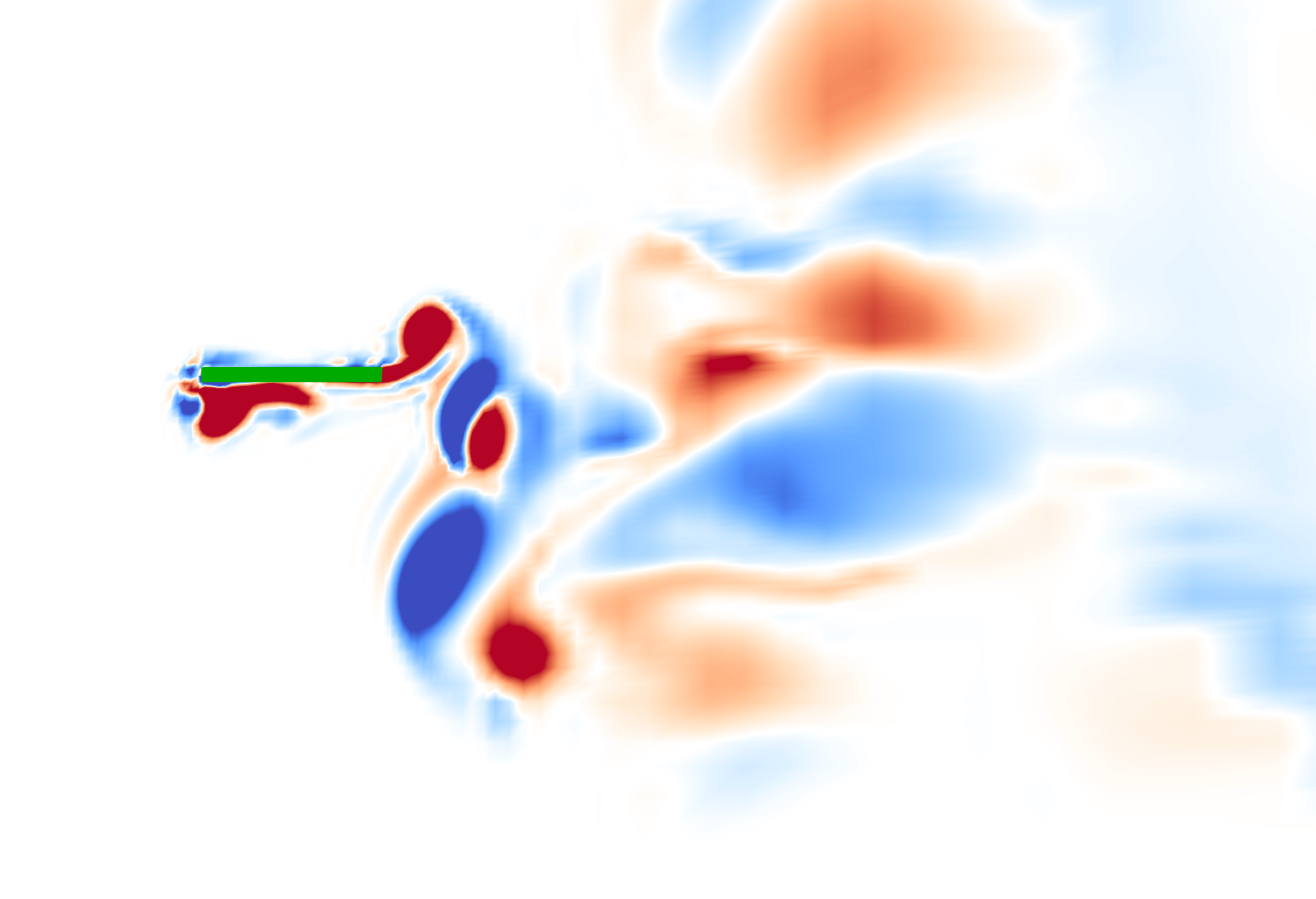}}
    \subfloat[Last control step (Baseline method)]{\includegraphics[width=0.33\textwidth]{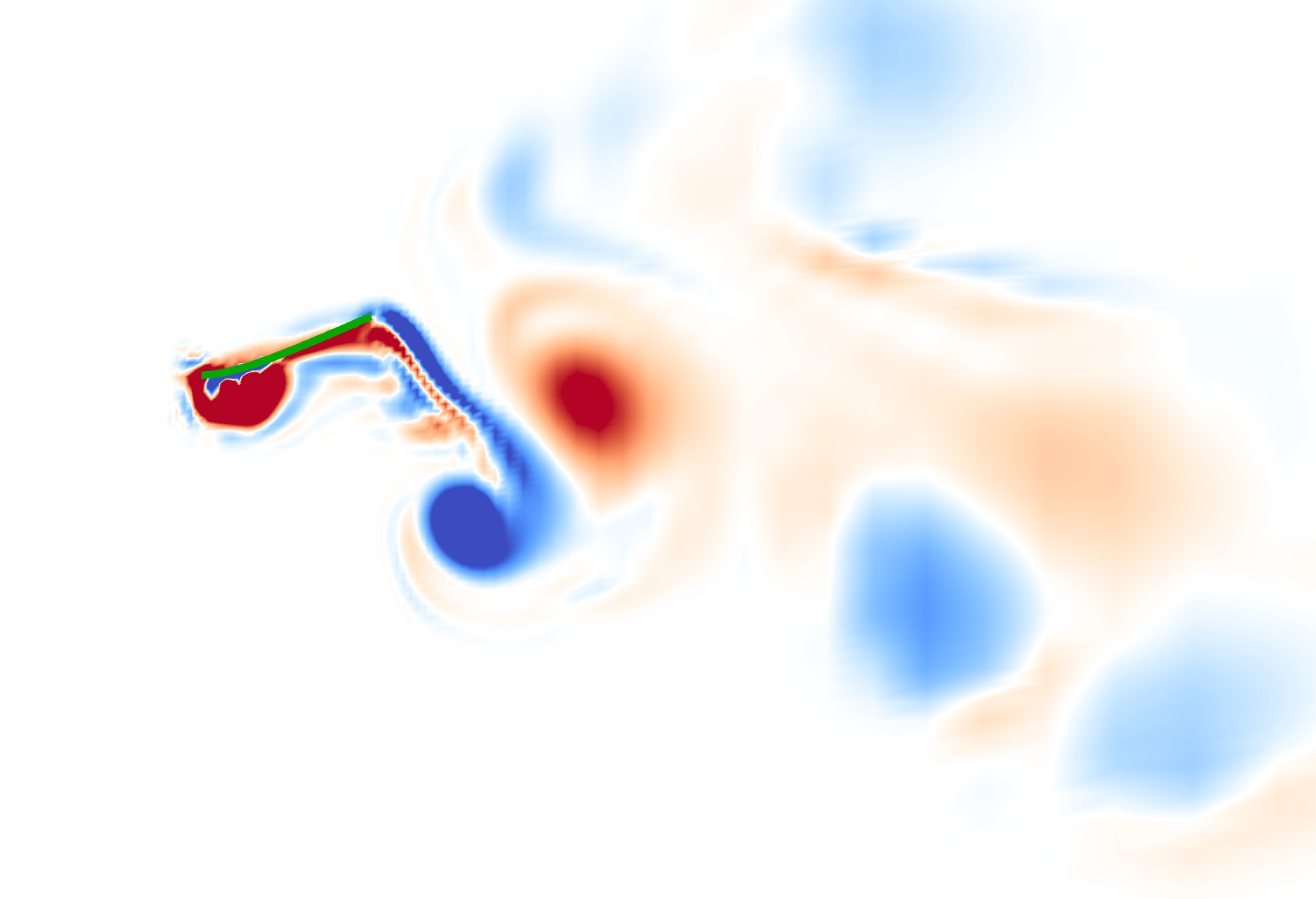}}    

    \vspace{-1.5ex}
    
    \subfloat{\includegraphics[width=0.36\textwidth]{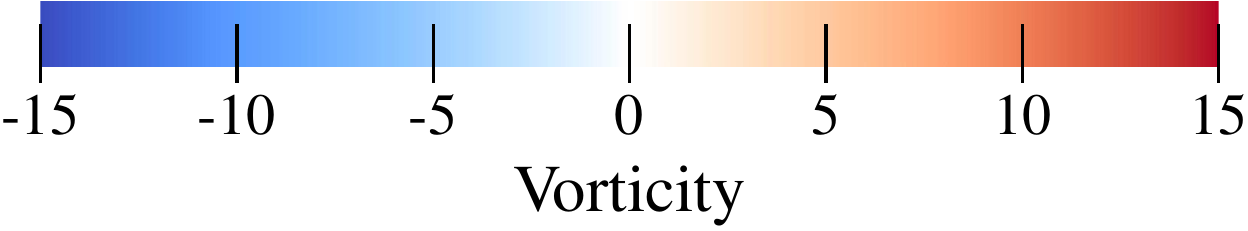}}

    \vspace{-2ex}
    \caption{The vorticity field in the maximize thrust efficiency case at the various control steps during the last 20 control steps ($i\in[60,80]$)(a-e) compared with the vorticity field at the last time step controlled by the baseline method (f). The position of fish-fin ray is indicated by (\boldgreenline)}
    
    \label{fig:MaxT:vorticity}
\end{figure}
Figure~\ref{fig:MaxT:vorticity} presents a sequence of vorticity fields captured at five representative DRL control steps within the last 20 steps of an episode, specifically at the 60th, 65th, 70th, 75th, and final control steps of an episode, as illustrated in panels (a) through (e). These frames reveal the intricacies of the fluid dynamics at play, capturing the heightened activity and interaction within the flow as a result of the DRL agent's control policy. When these fields are compared to the baseline method's output at the concluding step, shown in panel (f), the contrast is pronounced. The DRL agent's approach results in a vorticity field marked by a substantially increased number of vortices, which are arranged much closer together. This close arrangement indicates that the DRL agent effectively manages the spatial distribution of vortices, potentially translating to more effective thrust generation. The dense clustering of vortices may reflect a sophisticated control strategy that adeptly exploits fluid dynamics to optimize propulsion.

\begin{figure}[!htp]
    \centering
    \includegraphics[width=0.8\textwidth]{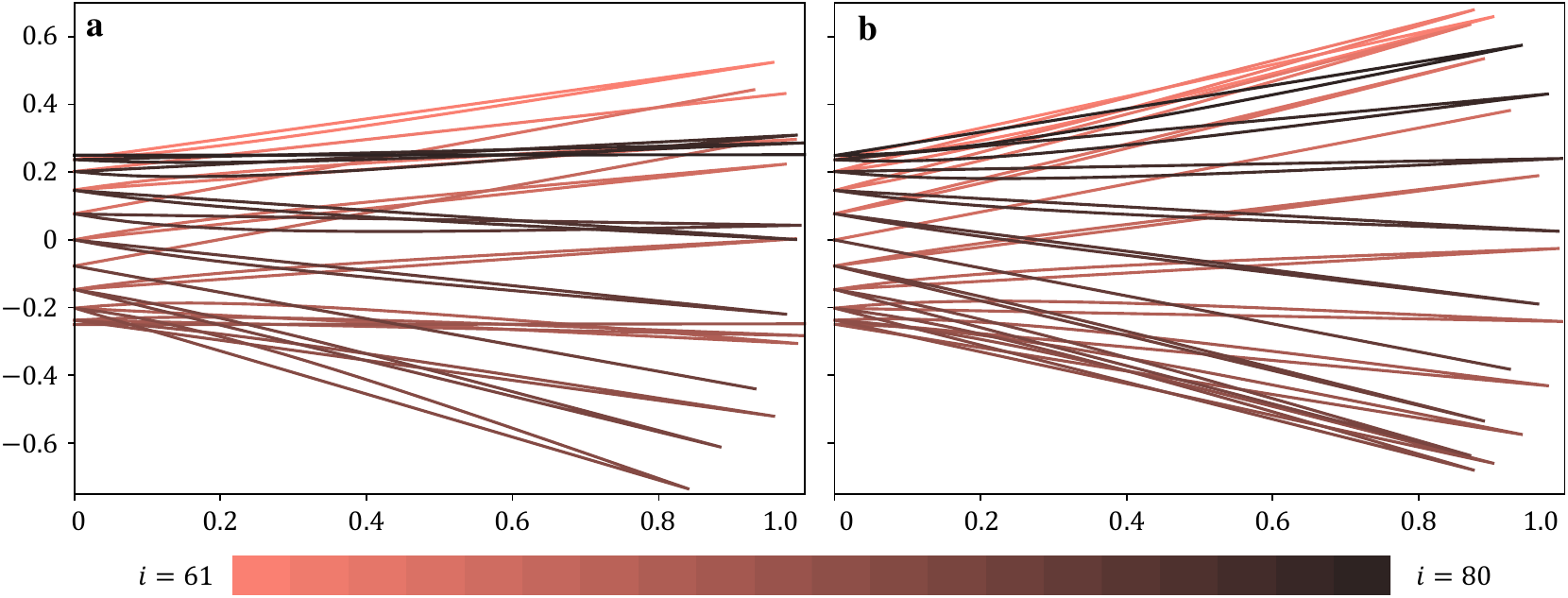}
    \caption{The shape and location trajectory of the fish-fin ray during the last 20 ($i\in[61,80]$ control steps, $i\in(61, 80]$), controlled by RL (a) compared with the baseline method (b). Darker colors indicate later control steps.}
    \label{fig:MaxT:traj}
\end{figure}

The shape and locations of the fin ray also reflects the distinction between the RL agent controlling and the baseline method. Fig.~\ref{fig:MaxT:traj}(a) shows the shape and locations of the fish fin ray within the last 20 control steps controlled by the RL agent. Compared with the baseline method (Fig.~\ref{fig:MaxT:traj}(b)), a more complex movement pattern of the fin ray is apparent. In particular, several more densely clustered regions of fin rays in adjacent control steps can be observed in the RL controlled episode compared to the baseline method, which again indicates the RL agent is able to generate more vortexes and potentially translating to more thrust.

\subsection{Maximize efficiency}
In the second case, the DRL agent is expected to find an control policy $\bm{\pi}^E$ which maximizes the overall propulsion efficiency $\eta$ in one episode. The optimization goal can be formulated as:
\begin{equation}
    \max_{a_i \sim \pi_{\theta^E}}\eta = \max_{a_i \sim \pi_{\theta^E}}\frac{F_T}{P} =
    \max_{a_i \sim \pi_{\theta^E}} \frac{\displaystyle\sum_{i=1}^{N}F_{T,i}}{\displaystyle\sum_{i=1}^{N} P_i}
    =\max_{a_i \sim \pi_{\theta^E}}\frac{\displaystyle\sum_{i=1}^{N} \int_{t_{i-1}}^{t_i} f_T\left(\bm{s}_i(t),\,a_i\right)\, \mathrm{d}t}{\displaystyle\sum_{i=1}^{N} \int_{t_{i-1}}^{t_i} p\left(\bm{s}_i(t),\,a_i\right) \, \mathrm{d}t}
    \label{eq:def_eff}
\end{equation}
where $p$ is the instantaneous power consumption introduced by taking the action $a_i$, while $P_i$ represents the accumulated power consumption in the $i^{th}$ control step and $P$ denotes the total power consumed by the RL agent in one episode. In contract to the first case, where the control objective can be easily formulated as a summation, the propulsion efficiency $\eta$ instead appears as a quotient of the \hot{total thrust} $F_T$ and total power consumption $P$, it is extremely challenging to accurately approximate the efficiency $\eta$ using a summation formula consisting of reward at each control step. To address this issue, we propose a training strategy called ``global searching and local Fine-tuning'' (GSLF, see detailed explanation in Section \ref{sec:dis:reward}), where the training of RL agent is divided into multiple stages and in each stage, different reward functions will be applied to approximate the control objective $\displaystyle\max_{a_i \sim \pi_{\theta^E}}\eta$.
In particular, here we divided the training into two stages and two different reward functions $r_{GS}$ and $r_{LF}$, are applied one by one. These two reward functions are calculated as:
\begin{equation}
    \label{eq:rpre}
    r_{GS\,;\, i} = c_3\frac{c_1 F_{T,i} - c_2 P_i}{c_1^2 + c_1 P_i} + c_4
\end{equation}
\begin{equation}
\label{eq:rfine}
    r_{LF\,;\, i} = F_{T,i} - P_i
\end{equation}
where $c_1 = 3\times10^4, c_2 = 4\times10^3$ are the hyperparameters related to the environment, $c_3 = 1000, c_4 = 1$ are the normalization parameters. 

\begin{figure}[!h]
    \centering
    \subfloat{\includegraphics[width=0.48\textwidth,]{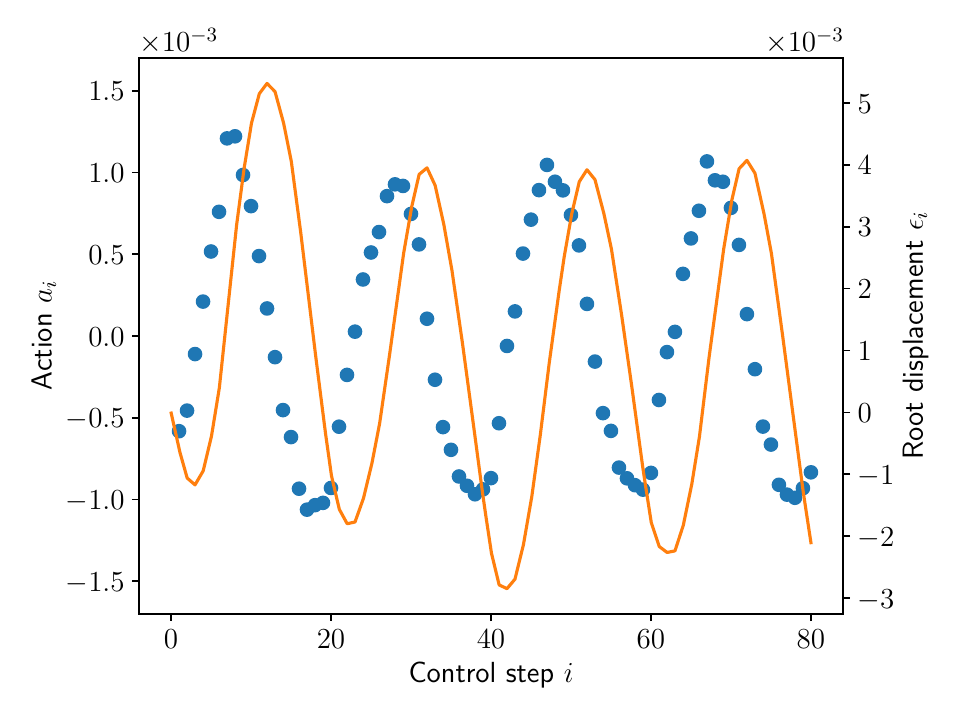}}
    \subfloat{\includegraphics[width=0.48\textwidth]{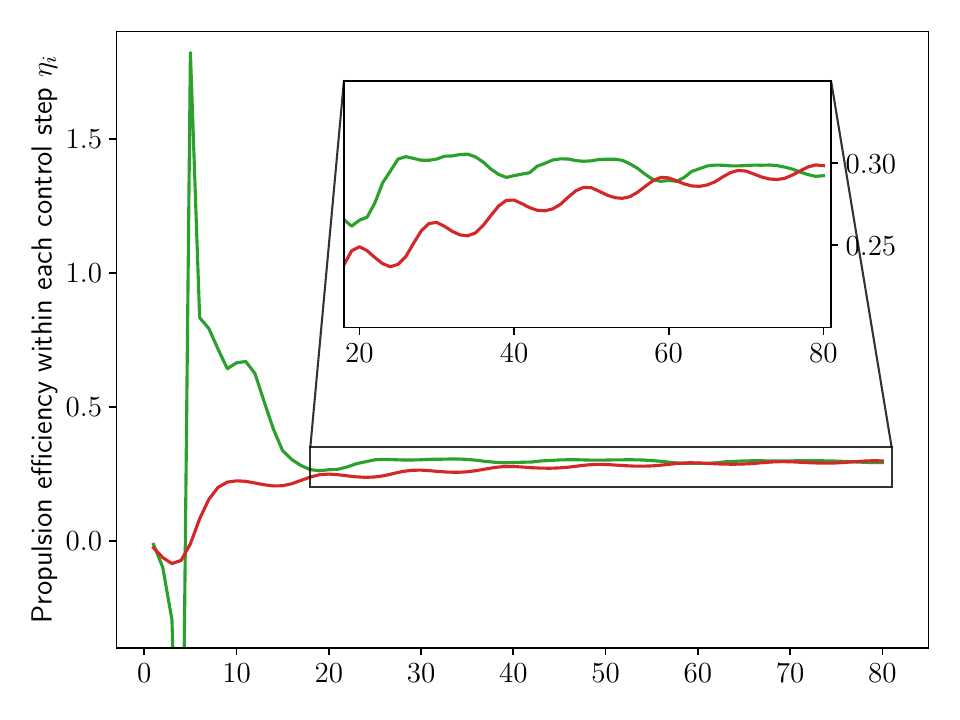}}
    \vspace{-12pt}
    \caption{Left panel: the root displacement $\varepsilon_i$ (\orangeline) and actions $a_i$ (\bluedot) learned by the RL agent.  Right panel: propulsion efficiency at each time step ($\eta_i$) controlled by RL agent (\greenline) compared with the highest efficiency control pattern found by the baseline method (\redline) during one episode.}
    \label{fig:MaxE:act}
\end{figure}

By applying the APT and the GSLF, the RL agent finds an optimal control policy, as depicted in figure \ref{fig:MaxE:act}. The left panel shows the control actions performed by the RL agent ($a_i$) as well as the root displacement ($\varepsilon_i$). 
Unlike the maximizing thrust case, where RL agent learns an aggressive control policy to max out thrust, here RL agent learns to take actions with moderate magnitudes with smooth transitions between different control steps. In particular, in the last 60 control steps ($i\in[21,80]$), a strong periodic pattern can be observed in the action ($a_i$) curve as well as the root displacement ($\varepsilon_i$) trajectory.This pattern, with a frequency closely matching that of the prescribed motions $\beta(t), h(t)$ (Eq.~\ref{eq:prescribedMotion}), suggests that the RL agent has learned to strategically adjust its actions to coordinate with the dynamic system. This alignment of frequencies indicates the agent's capability to enhance propulsion efficiency by responding effectively to its surrounding environment. While during the first 20 steps ($i\in[1,20]$), the environment is transitioning from a stationary state (i.e. initial condition) to a more periodic state introduced by the motion of the fin ray. Such complex transition behaviour is reflected in the right panel of Fig.~\ref{fig:MaxE:act} which shows the history of propulsion efficiency at each control step: $\eta_i$, which is defined as 
\begin{equation}
    \eta_i \doteq \frac{\sum_{j=1}^i F_{T,j}}{\sum_{j=1}^i P_j}
\end{equation} 
During the transitional stage (i.e. $i\in[1,20]$), RL-controlled episode has significantly higher propulsion efficiency $\eta_i$ compared to the baseline method and for most the control steps, RL controlled episode maintains a higher efficiency. Only in the last few steps, does the baseline method achieve a slightly higher efficiency. The distinctive efficiency difference indicates the RL agent is significantly better in controlling a complex transitional dynamic system compared to the baseline method. Although at the last few steps, the propulsion efficiency of RL controlled episode is surpassed by the baseline method, resulting in a slightly lower overall efficiency ($29.23\%$ by RL compared to $29.84\%$ by baseline method), the DRL controlling is expected to perform significantly better than the baseline method because the DRL does not rely on any prior knowledge. However, in the baseline method, we enforce the frequency of the root displacement $\varepsilon(t)$ to be exactly the same value of the frequency of the prescribed motions, which is impossible to achieve in real-world experiments where the other motions of the fin ray cannot be precisely measured/enforced. Besides, RL agent is only trained to maximize the propulsion efficiency in a certain number of control steps and RL has shown promising performance in the transitional stage, which take a significant portion of the overall episode. If the RL is trained to control a longer episode where the transitional stage takes smaller ratio, we believe DRL will achieve higher efficiency in the periodic stage.

\begin{figure}[!ht]
    \centering
    \subfloat[The $60^{th}$ control step (DRL)]{\includegraphics[width=0.33\textwidth]{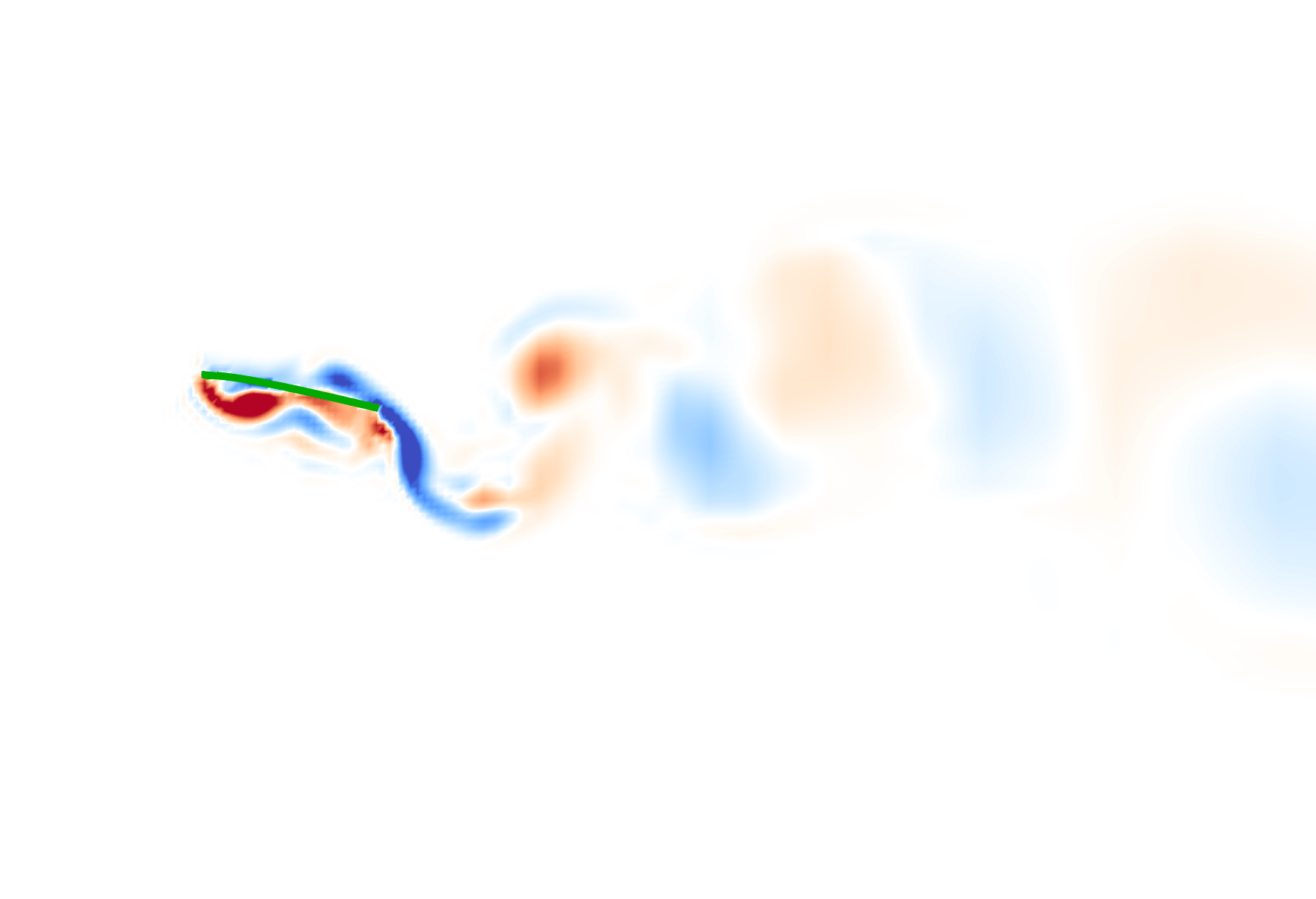}}
    \subfloat[The $65^{th}$ control step (DRL)]{\includegraphics[width=0.33\textwidth]{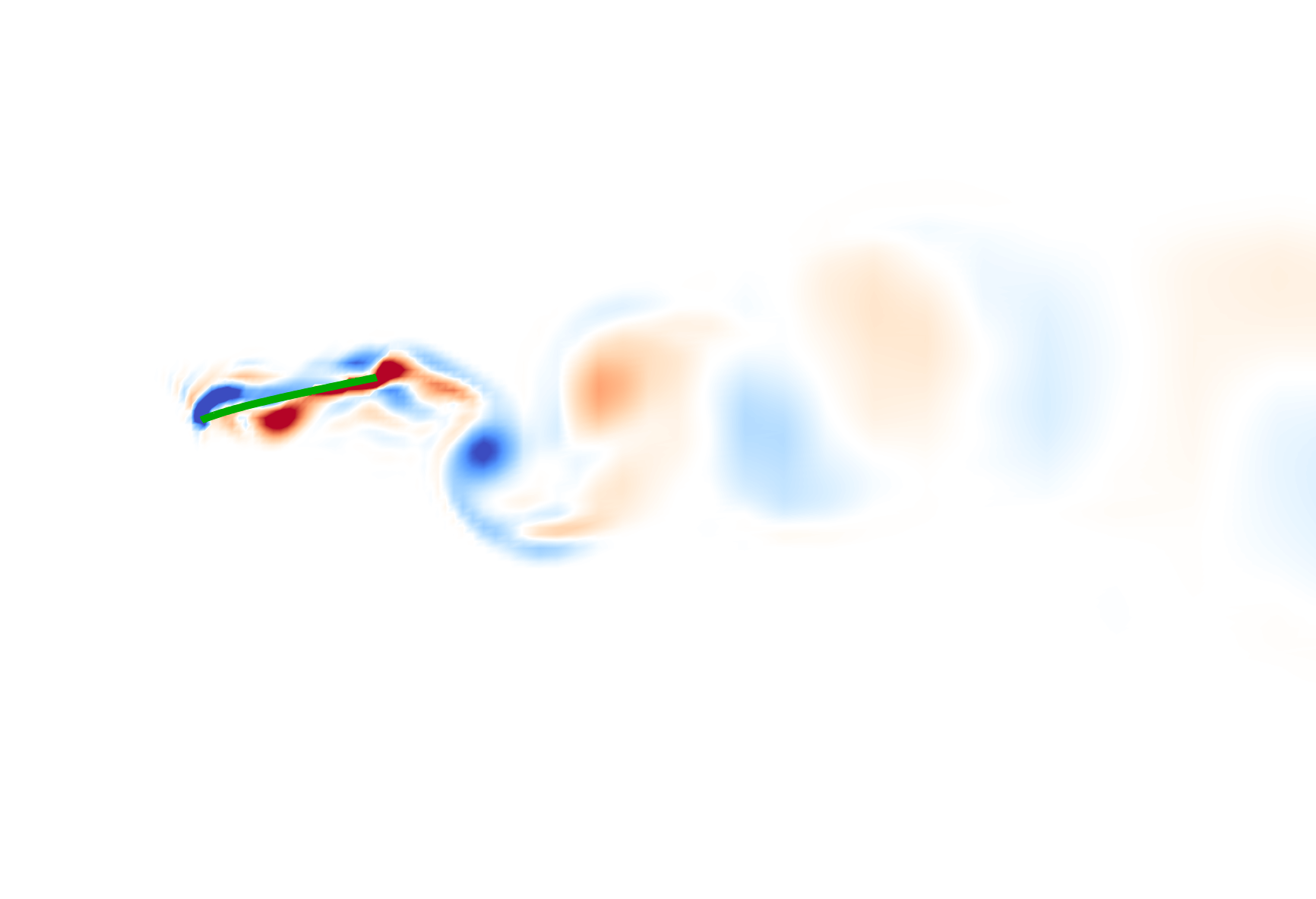}}
    \subfloat[The $70^{th}$ control step (DRL)]{\includegraphics[width=0.33\textwidth]{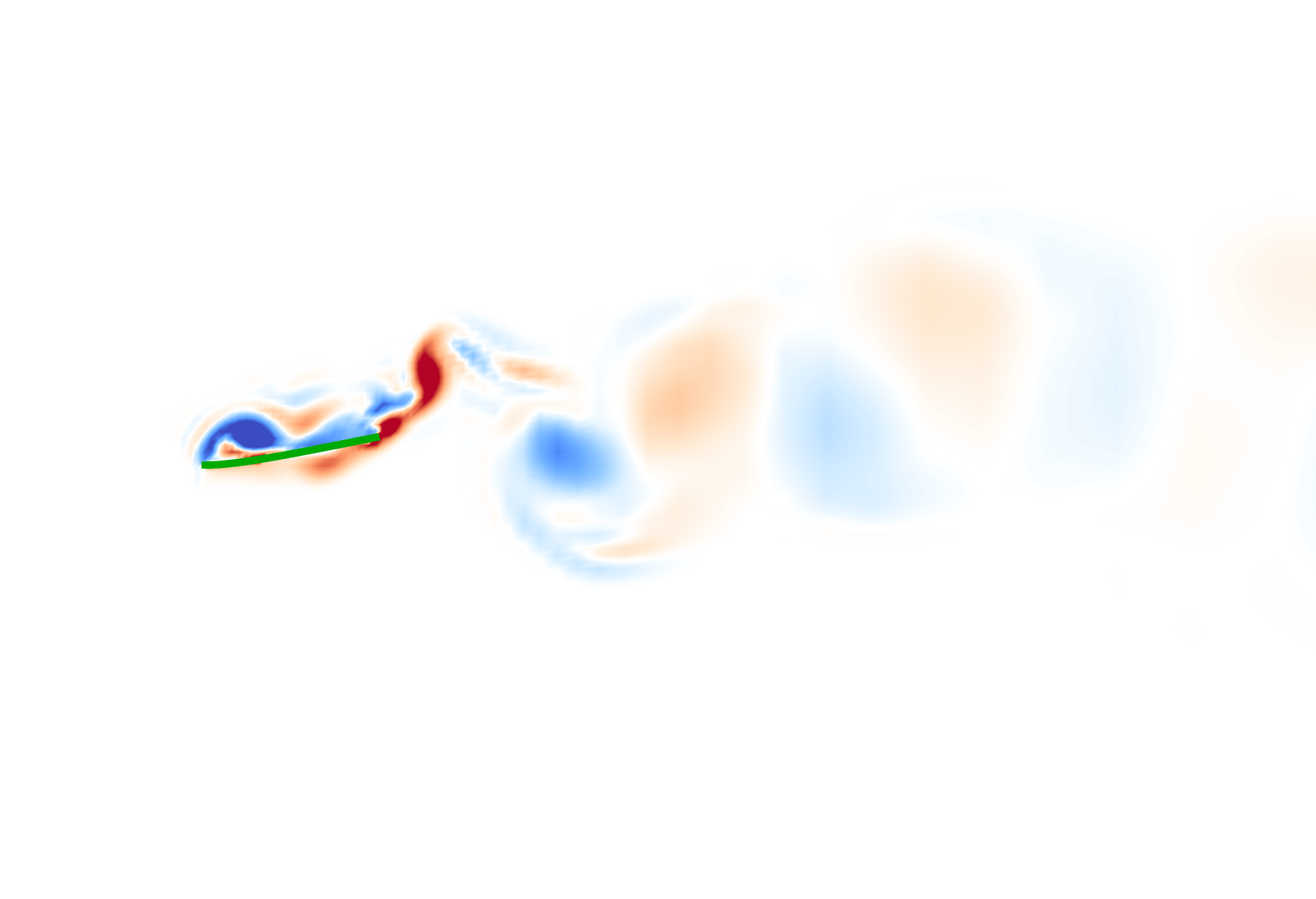}}\\

    \vspace{-2ex}

    \subfloat[The $75^{th}$ control step (DRL)]{\includegraphics[width=0.33\textwidth]{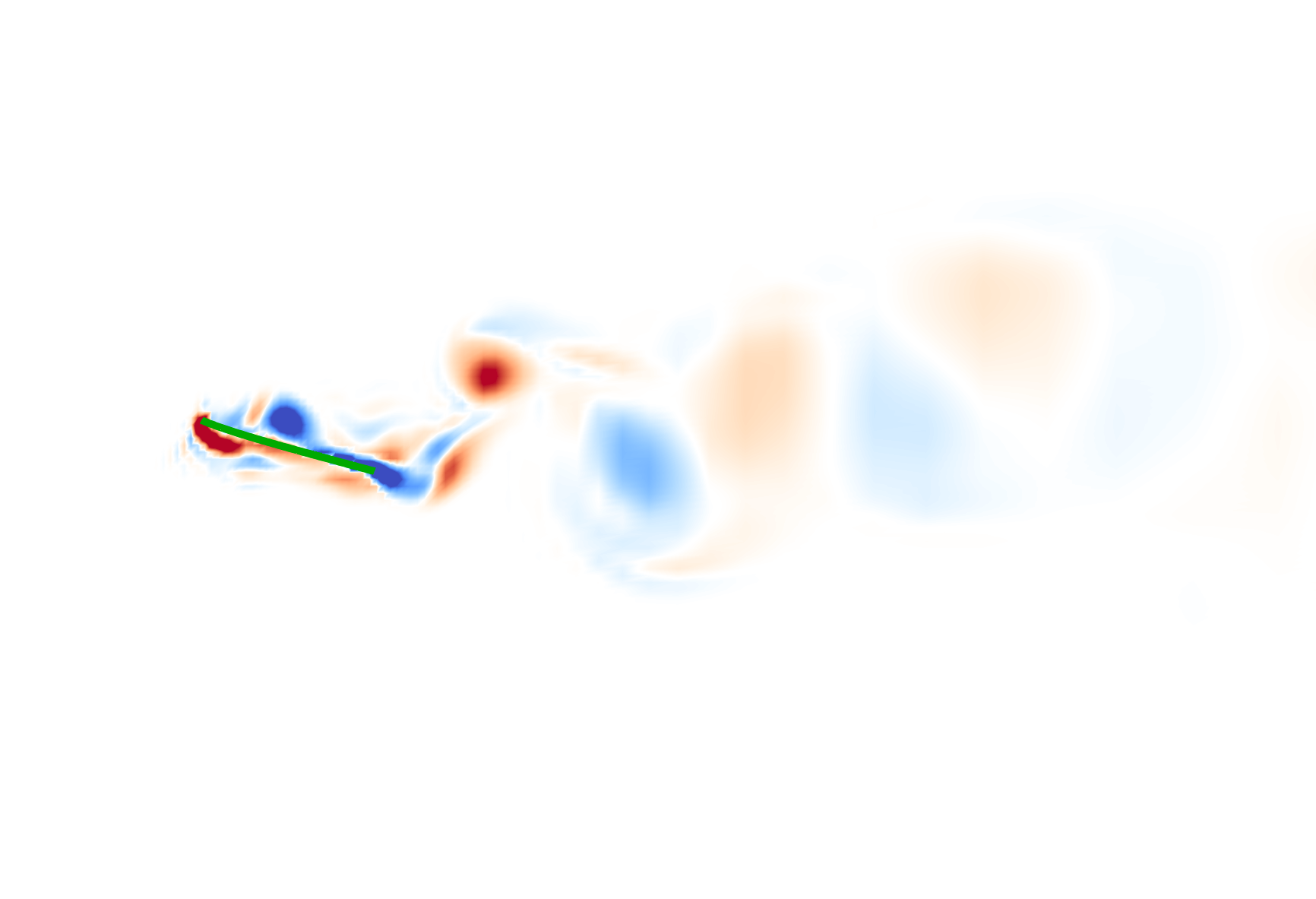}}
    \subfloat[Last control step (DRL)]{\includegraphics[width=0.33\textwidth]{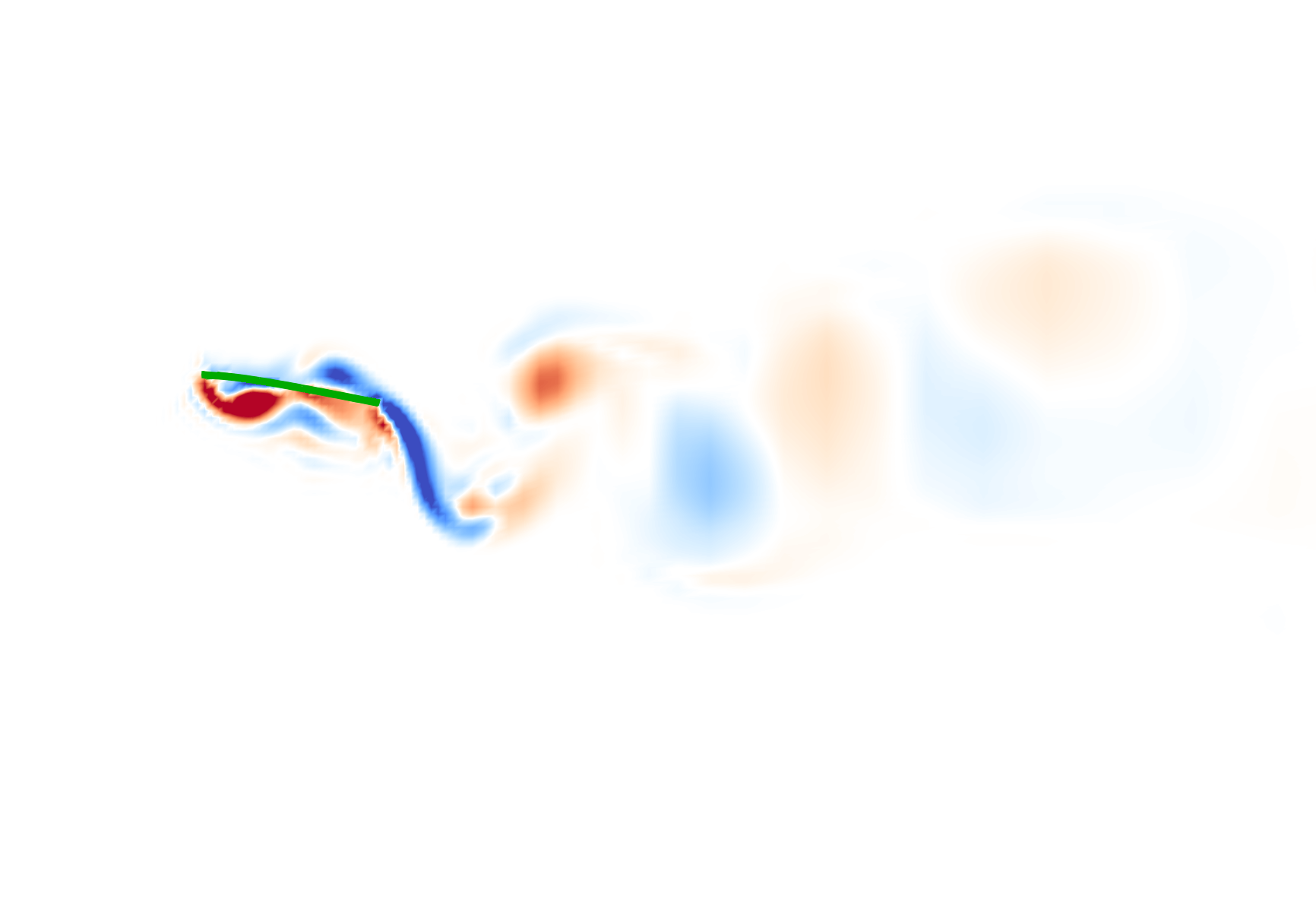}}
    \subfloat[Last control step (Baseline method)]{\includegraphics[width=0.33\textwidth]{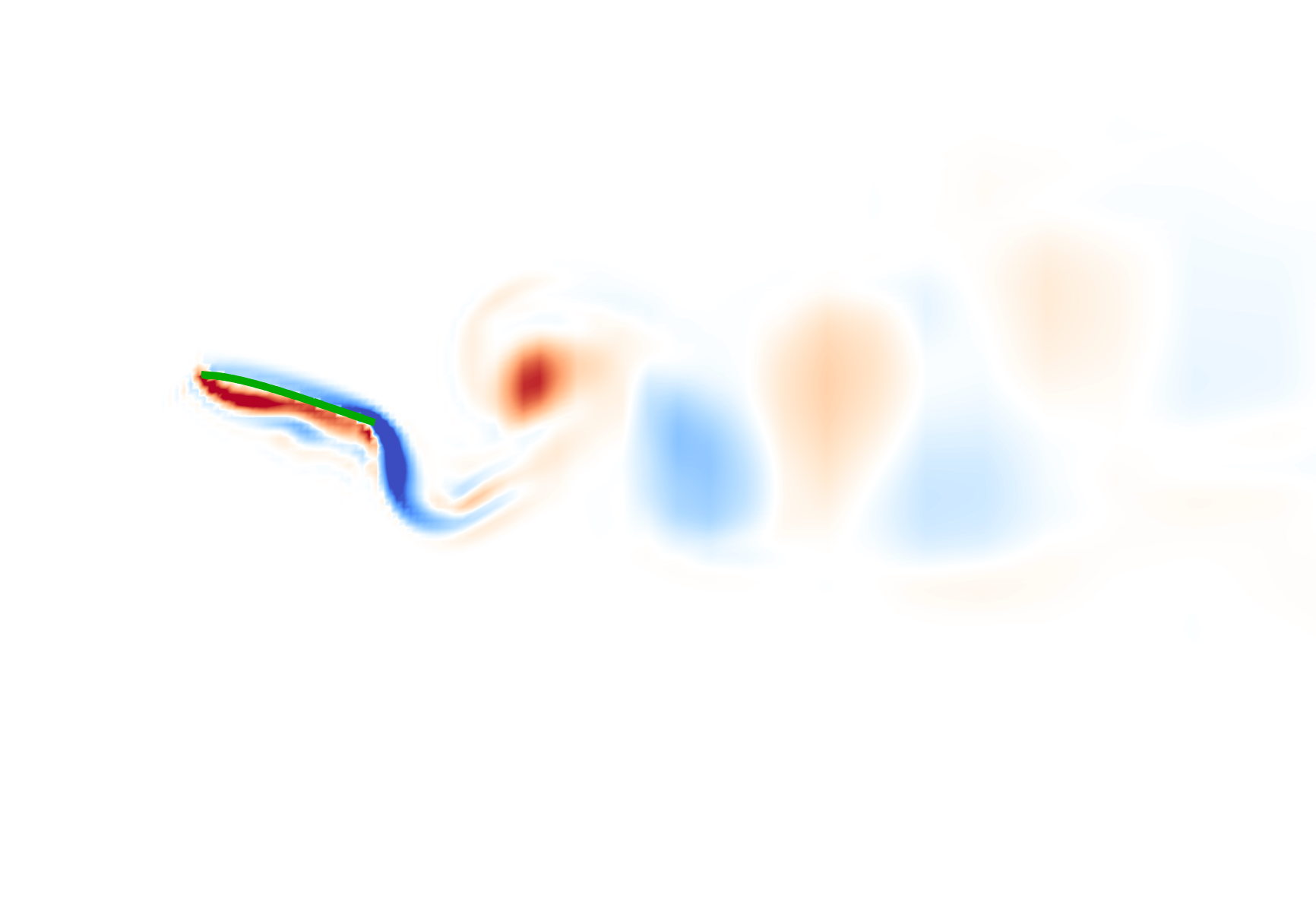}}

    \vspace{-1.5ex}
    
    \subfloat{\includegraphics[width=0.36\textwidth]{figs/colorbarContour.pdf}}

    \vspace{-2ex}
    \caption{The vorticity field in the maximize propulsion efficiency case at the various control steps during the last 20 control steps ($i\in[60,80]$)(a-e) compared with the vorticity field at the last time step controlled by the baseline method (f). The position of fish-fin ray is indicated by (\greenline)}
    
    \label{fig:MaxE:vorticity}
\end{figure}

Figure~\ref{fig:MaxE:vorticity}(a-e) presents a sequence of vorticity fields captured at five representative DRL control steps within the last 20 steps of an episode, specifically at the 60th, 65th, 70th, 75th, and the final control step of an episode. These frames reveal the flow field influenced by the DRL controlled fish fin ray. When these fields are compared to the baseline method's output at the concluding step, shown in Fig.~\ref{fig:MaxE:vorticity}(f), the DRL agent's approach results in a similar vorticity field, which indicates the DRL successfully learns to leverage the dynamics of the fluid environment by adopting a sinusoidal-like control policy that shares similar frequency as the prescribed motions. The similarity in the distribution of vortices may reflect a sophisticated control strategy that adeptly exploits fluid dynamics to optimize propulsion efficiency.

\begin{figure}[!htp]
    \centering
    \includegraphics[width=0.8\textwidth]{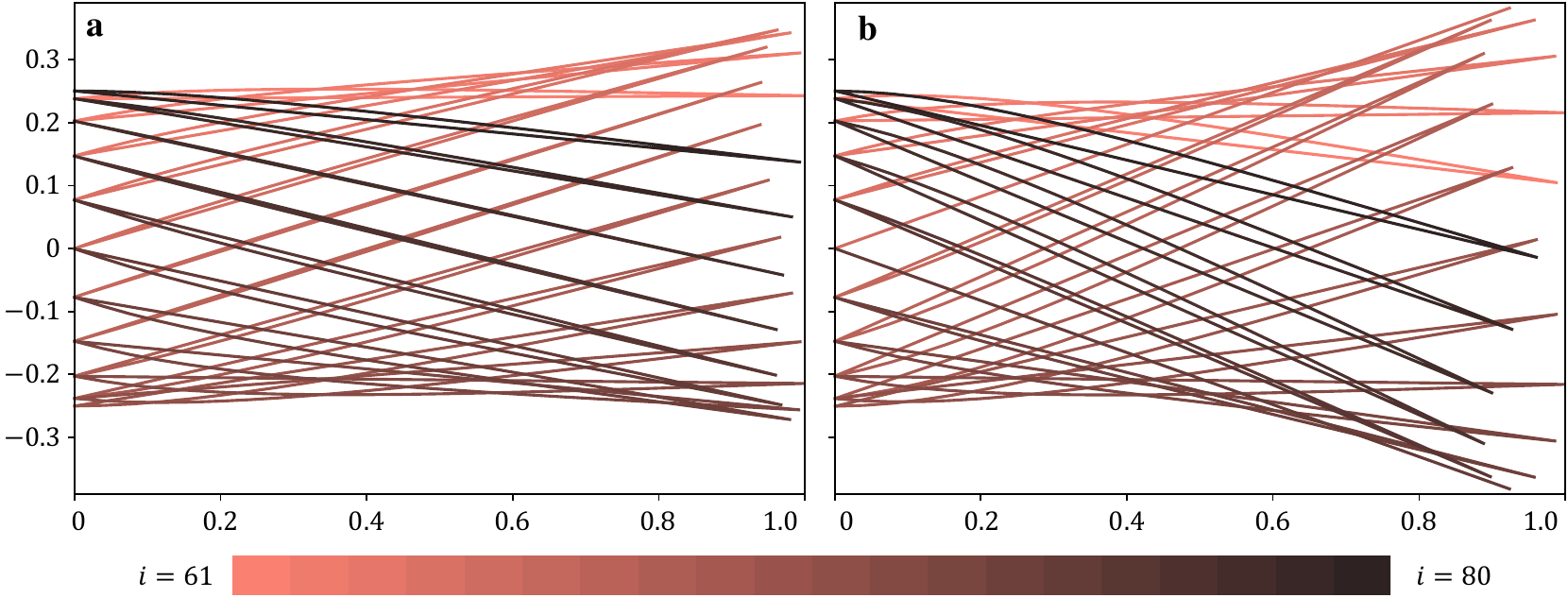}
    \caption{The shape and location of the fish-fin ray during the last 20 steps control steps, $i\in[61, 80]$), controlled by RL(a) and the baseline method (b). Each line represents the shape and location of the fin ray in one control step.}
    \label{fig:MaxE:traj}
\end{figure}
The similarity between the RL controlled episode and the baseline method in the maximize efficiency case can be further verified by the figure \ref{fig:MaxE:traj}, which depicts the shape and location of the fish fin ray in the last 20 control steps. Both the RL controlled fin ray (Fig.~\ref{fig:MaxE:traj})(a)) and the baseline method controlled fin ray (Fig.~\ref{fig:MaxE:traj}(b)) share a similar range of the amplitude of the trailing-edge of the fin ray. Although the overall visual similarities, small distinctions between the RL controlled fin ray and the baseline method controlled fin ray can still be observed. In particular, the baseline method controlled fin ray shows a strictly symmetric pattern about $y=0$. However, the symmetry property is not strictly satisfied in the RL controlled fin ray. Such not completely symmetric pattern also explains the slightly lower efficiency RL agent achieved in the last few control steps. Such asymmetric pattern is also related to the transitional stage where RL agent effectively improve the efficiency compared to baseline method. When compared to the maximizing thrust case (Fig.~\ref{fig:MaxE:traj}(a)), RL agent controlled fin ray shows a significantly different pattern, with the trailing edge distribution range shrunken by half, indicating the RL agent effectively learns different strategies to achieve different control objectives. 

\section{Discussion}
\label{sec:dis}

\subsection{Comparative Efficacy of APT with Conventional DRL Training Strategies} 
Having demonstrated APT's capability in handling complex FSI problems, we now present a comparative analysis to underscore its advantages. In this section, we compare APT against two conventional RL training strategies: Single Environment Training (ST) and Synchronous Parallel Training (SPT), highlighting the superior sample efficiency and training speed offered by APT.

\subsubsection{Testing Environment for Benchmarking}
Direct interaction with high-fidelity FSI simulations is computationally prohibitive for conventional DRL training methods. To facilitate a fair comparison, we employ a one-dimensional chaotic system governed by the Kuramoto-Sivashinsky (KS) equation as a test environment. The KS system is often used as a model problem for turbulence study due to its chaotic behavior~\cite{cvitanovic2010state}. Here, the KS environment is controlled by four actuators equally-distributed in space aimed at minimizing energy dissipation and total power input. The governing dynamics are expressed as,
\begin{equation}
u_t + u_{xx} + u_{xxxx} + u u_x = f(x,t), \quad x \in [0,l], , t \in [0,+\infty],
\end{equation}
where $u$ is the state variable, and $f$ represents the actuator-induced source term. The source term is modeled as a sum of Gaussian functions centered at the actuator locations, 
\begin{equation}
    f(x,t) = \sum_{i=1}^{4}\frac{a_i(t) e^{-(x-x_i)^2/2}}{\sqrt{2\pi}},
\end{equation}
where $x_i \in \{0,\,l/4,\,l/2,\,3l/4\}$ is the spatial coordinates of the actuators, and $\abm = \{a_i(t)\}_{i=1,2,3,4} \in [-0.5,0.5]^4$ is the control parameters. To minimize the energy dissipation of the system with minimum input power, the reward function is designed as follows,
\begin{equation}
    r = -\frac{1}{T l}\int_{t_0}^{t_0+T}\int_0^l \left((\frac{\partial^2 u}{\partial x^2})^2 +(\frac{\partial u}{\partial x})^2 + u f\right)\,dx\, dt 
\end{equation}
where $T$ denotes the duration of one control step. The environment is simulated numerically using the finite difference method, where the convection term is discretized by the second-order upwind scheme, and the second and fourth derivatives are discretized by the $6^{th}$ order central difference scheme. The $4^{th}$ order Runge-Kutta scheme is used for time integration with a timestep of 0.001 within a spatial domain of $l=8\pi$ discretized into 64 grid points.

\subsubsection{Benchmarking Results and Insights}

\begin{figure}[!htp]
    \centering
    \includegraphics[width=\textwidth]{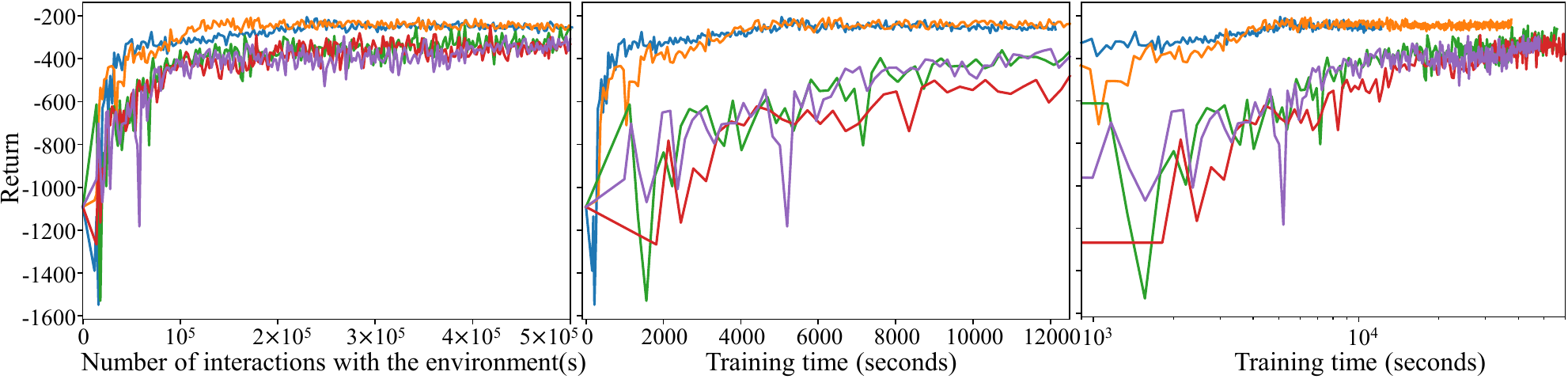}
    \caption{Performance and training time analysis across different DRL training strategies in the KS environment. The left panel illustrates the performance curves for APT with 8 parallel environments (\blueline) and four parallel environments (\orangeline), compared with the performance curves for ST (\purpleline) and SPT with 8 (\greenline) and 4 (\redline) parallel environments. The middle panel presents a linear scale comparison of the training time required by each method, while the right panel offers a logarithmic scale perspective, enhancing the visibility of differences in the later stages of training.}
    \label{fig:aprl_compare}
\end{figure}

Figure~\ref{fig:aprl_compare} compares the performance of APT, ST, and SPT training strategies within the SAC framework in the KS environment. The performance curves clearly demonstrate APT's superior efficiency in sample utilization and speed of convergence. When running four parallel environment  (APT-4 ~\orangeline), it exhibits remarkable sample efficiency, achieving optimal policy convergence with fewer than $1.5\times10^5$ total interactions with the environment. In contrast, the conventional DRL training strategies (ST and SPT) cannot achieve the optimal policy even after $5\times10^5$.
When utilizing eight parallel environments, the initial phase of APT-8 (\blueline) reveals a quicker ascent in total return, attributed to the increased data availability from the higher number of parallel environments. However, this benefit is transient, as the APT-8 eventually shows a slightly diminished sample efficiency, necessitating under $2.5\times10^5$ interactions for convergence. This is because a surplus of parallel environments tends to saturate the replay buffer with outdated data, inadvertently hampering overall sample efficiency. This phenomenon is also observed with the traditional DRL training methods, where scaling up the number of environments in parallel fails to notably enhance sample efficiency. This pattern suggests that the quantity of samples is not the primary constraint; rather, the critical factor impeding the training efficiency of conventional DRL strategies is the suboptimal utilization of the accumulated interaction experiences.

The advantage of APT becomes even more pronounced when examining the training speed. As depicted in the middle panel of Figure~\ref{fig:aprl_compare}, both APT-4 and APT-8 configurations showcase a rapid initial increase and converge to the optimal policy in $5\times10^3$ seconds, while , ST and SPT require more than an order of magnitude ($>5\times10^4$ seconds) to achieve comparable levels of performance, as further detailed in the logarithmic scale of the right panel. The introduction of more environments in parallel can only have marginal gains in terms of training speed for these conventional methods. In contrast, APT's asynchronous architecture significantly bolsters both the training speed and sample efficiency by effectively leveraging the already collected dataset.

\subsection{Reward formulation for non-additive control objectives}
\label{sec:dis:reward}
RL intrinsically depends on additive reward functions, yet many control objectives, such as efficiency, are inherently non-additive. For example, efficiency is usually defined as a quotient instead of a summation. This discrepancy necessitates the transformation of non-additive goals into additive reward functions that peak at the same global optimum within the state-action space as the original objective. Identifying such reward functions is often challenging, particularly when the location of the global optimum is unknown, which is typical in all RL scenarios. In this section, we discuss how we design additive reward functions that approximate the global optimum for maximizing propulsion efficiency, which is non-additive.

\subsubsection{Transitioning non-addable goals to additive rewards}
A straightforward additive approximation of efficiency, as defined in discrete terms (see Eq. \ref{eq:def_eff}), is,
\begin{equation}
\widetilde{\eta} = \sum_{i=1}^{N} \frac{{F_{T,i}}}{P{i}},
\label{eq:def_eff:rubbish}
\end{equation}
yet this form is not an ideal reward function. Apart from failing to align its global maximum with that of the true efficiency $\eta$, it suffers from instability, particularly when power consumption $|P_i|$ is minimal. Considering the possible negativity of $P_i$ in our study, which indicates energy released from the fish fin ray, directly adopting Eq. \ref{eq:def_eff:rubbish} as a reward function would severely hinder the convergence of DRL training.

To find a stable and accurate approximation, we propose a reward function that captures the incremental change in efficiency caused by each action the DRL agent takes. Accordingly, we derive the following reward function,
\begin{equation}
\begin{split}
    r_j = \eta_{j} - \eta_{j-1} 
         &= \frac{\displaystyle\sum_{i=1}^{j} F_{T,i}}{\displaystyle\sum_{i=1}^{j} P_i} - \frac{\displaystyle\sum_{i=1}^{j-1} F_{T,i}}{\displaystyle\sum_{i=1}^{j-1} P_i}
           = \frac{
                    \left( F_{T,j} + \displaystyle\sum_{i=1}^{j-1}  F_{T,i}\right) 
                    \displaystyle\sum_{i=1}^{j-1}  P_i
                    - 
                    \left( P_j + \displaystyle\sum_{i=1}^{j-1} P_i\right)
                    \displaystyle\sum_{i=1}^{j-1} F_{T,i}
                }
                {
                    \left(P_j + \displaystyle\sum_{i=1}^{j-1} P_{i}\right)
                    \displaystyle\sum_{i=1}^{j-1} P_i 
                }\\
            &= \frac{
                    F_{T, j}\displaystyle\sum_{i=1}^{j-1} P_{i}
                    -
                    P_{j} \displaystyle\sum_{i=1}^{j-1} F_{T,i}
                }
                {
                    \left(P_{j} + \displaystyle\sum_{i=1}^{j-1} P_{i}\right)
                    \displaystyle\sum_{i=1}^{j-1} P_{i} 
                }
\end{split}
\label{eq:converting}
\end{equation}
Assuming an infinitely long episode allows us to treat \hot{cumulative thrust} $\displaystyle\sum_{i=1}^{j-1} F_{T,i}$ and cumulative power $\displaystyle\sum_{i=1}^{j-1} P_{i}$ as constants. This assumption simplifies Eq.~\ref{eq:converting} to the following form,
\begin{equation}
    r_i \approx \frac{c_1 F_{T,i} - c_2 P_i}{c_1^2 + c_1 P_i},
\end{equation}
where $c_1$, $c_2$ are the constants, representing typical values of thrust/power generated/consumed over an entire episode. For practical training, we introduce normalization constants $c_3$ and $c_4$, leading to a reward function as shown in Eq.~\ref{eq:rpre}. This adaptation ensures stability and alignment with the global maximum of the original control objective.

\subsubsection{Global searching and local Fine-tuning (GSLF)}
Finding a universally applicable additive alternative for efficiency $\eta$ is very challenging; however, obtaining localized approximations for various regimes is more achievable. In this work, we introduce a global searching and local fine-tuning (GSLF) algorithm using an set $\mathcal{R}$ of two reward functions $r_{GS}$ and $r_{LF}$ for optimizing the efficiency $\eta$. Note that the GSLF method is adaptable, capable of incorporating any number of functions sequentially applied during training,
\begin{equation}
    \mathcal{R} = \left\{r_{s_1}\left(\bm{S}_1,\bm{A}_1\right),\, r_{s_2}\left(\bm{S}_2,\bm{A}_2\right),\dots \,r_{s_n}\left(\bm{S}_n,\bm{A}_n\right)\right\}, \quad \bm{\Omega}_i = \{\bm{S}_i,\bm{A}_i\} 
\end{equation}
where $r_{s_i}$ represents the $i^{th}$ reward function used to train the agent; $\bm{S}_i$ and $\bm{A}_i$ represent the states and actions spaces consists of the bunch of trajectories evaluated by the function $r_{s_i}$, while $\bm{\Omega}_i$ is the state-action space consisted of $\bm{S}_i$ and $\bm{A}_i$. Each reward function, $r_{s_i}$, is a ``good'' approximator within a specific region of the state-action space $\bm{\Omega}_i$, avoiding to search for a global approximation. The selection of reward functions follows a strategic sequence: the initial reward should enable stable optimization at a global scale, while subsequent functions should increase in accuracy and specificity around the optimal policy within a narrowing state-action space. This staged strategy uses initial rewards for global exploration (global searching), directing the search towards promising regions that may contain the optimal policy, and later rewards for precise optimization within these high-reward zones (local fine-tuning).

The ``global searching'' reward functions need not perfectly match the global optimum location of the original control goal within the state-action space, but their greater stability and satisfactory global approximation help to limit the DRL agent's exploration to a smaller, high-reward area. Conversely, ``local fine-tuning'' functions may be unstable outside their intended high-reward region or exhibit distinct global maxima; nonetheless, they effectively pinpoint the optimal policy within a confined space or further restrict the search for subsequent reward functions. Ideally, each subsequent state-action space is nested within its precursor ($\bm{\Omega}_i \subset \bm{\Omega}_{i-1}$), though in practice, exploration may extend beyond prior bounds ($\{\bm{S}_i,\,\bm{A}_i\} \not\subset \bm{\Omega}_{i-1}$), albeit within a significantly reduced dimensional scope  ($d_{\bm{\Omega}_i}$ < $d_{\bm{\Omega}_{i-1}}$).

\begin{figure}[!ht]
    \centering
    \includegraphics[width=\textwidth]{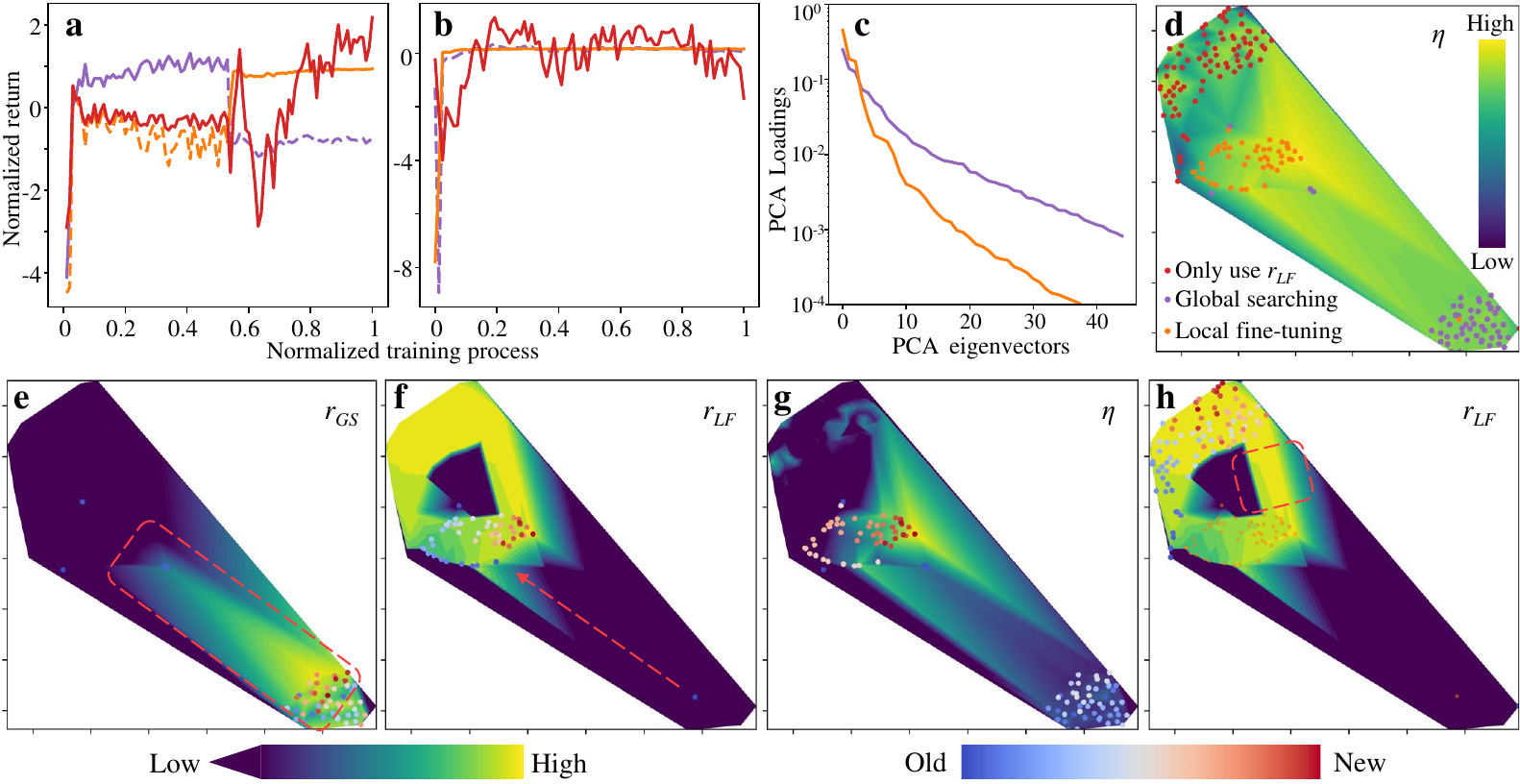}
    \caption{Analysis of GSLF method in the case of maximizing propulsion efficiency $\eta$. (a-b) Normalized return during the training process based on three different reward functions: $r_{GS}$ (\purpleline), $r_{LF}$ (\orangeline) and the efficiency $\eta$ (\redline). The dashed parts of $r_{GS}$ and $r_{LF}$ indicate the reward function is only used to evaluate, while the solid parts represent the reward function is used for training the RL agent. (c) The weight of the first 45 principle components of the state \& action space of the testing trajectories during the training process at two stages: pre-training stage (\purpleline) and fine-tuning stage (\orangeline). (d) The distribution of all the testing trajectories chosen during training in the state \& action space. The state \& action space is projected to a two-dimensional space for visualization based on t-distributed stochastic neighbor embedding (t-SNE) method. The contour is colored based on the efficiency $\eta$. (e-h) The testing trajectories (dots) projected to the two dimensional t-SNE space. The contour is colored by the normalized reward function $r_{GS}$ (e), $r_{LF}$ (f), the efficiency $\eta$ (g) and $r_{LF}$ (h), respectively. The color range of the contours are truncated to high reward regions. The testing trajectories (dots) are colored by the order of the RL agent choose these testing trajectories. In the panel (h), trajectories from $\bm{\Omega}_{LF}$ (represented by orange dots \orangedot) are added for comparison. }
    \label{fig:reward-tsne}
\end{figure}

Practically, we deploy GSLF using two chosen reward functions, $r_{GS}$ for global exploration and $r_{LF}$ for local optimization, in  in training an RL agent to discover the optimal policy ($\bm{\pi}_E$) for maximum propulsion efficiency. Here, we use the data collected from the DRL training on a shorter episode containing $20$ control steps for a better visualization. Figure~\ref{fig:reward-tsne}(a-b) depicts the transition from $r_{GS}$ to $r_{LF}$, reflecting a progression from a global, exploratory search to a local, efficiency-optimizing fine-tuning.
Fig.~\ref{fig:reward-tsne}(a) shows the performance curve during training. During the first half of the training process, the agent was trained with $r_{GS}$, followed by $r_{LF}$ in the latter half. Although the return demonstrated consistent growth throughout the early training stage, the actual propulsion efficiency declined, suggesting that the learned policy was confined within a suboptimal region of the state-action space $\bm{\Omega}_{GS}$. This region was characterized by a significant deviation of the maximal efficiency determined by $r_{GS}$ compared to the true maximum efficiency $\eta$. However, upon initiating the fine-tuning phase with $r_{LF}$, a notable surge in propulsion efficiency $\eta$ ensued, as $r_{LF}$ continued to climb gradually. Notably, the return calculated with $r_{GS}$ maintained a degree of stability across both stages, despite a slight decrease during fine-tuning. This contrasted with the return as measured by $r_{LF}$ during the initial global search phase, where notable variability highlighted underscoring its inappropriateness for the initial training stage. However, in the fine-tuning stage, $r_{LF}$ showcased impressive stability, providing effective guidance towards the optimal policy. It is evident that $r_{LF}$ offers a closer representation of $\eta$ in proximity to the optimal policy found in $\bm{\Omega}_{LF}$. Yet, completely skipping the global search and exclusively relying on $r_{LF}$ for training is not feasible, as reflected by the Fig.~\ref{fig:reward-tsne}(b), where the curve illustrates that the training conducted solely with $r_{LF}$ yielded negligible improvements in efficiency $\eta$ and was marked by considerable fluctuations, notwithstanding the consistent return as assessed by $r_{LF}$. This pattern suggests the policy is entrapped within a local maximum of the uniquely $r_{LF}$-defined state-action space $\bm{\Omega}'_{LF}$, which substantially diverges from the true optimal policy.

The effectiveness of GSLF can be further demonstrated by Fig.~\ref{fig:reward-tsne}(c-h). Figure~\ref{fig:reward-tsne}(c) shows the weights of the first 45 principal components of the test state-action trajectories during the global searching and local fine-tuning stages using the Principal component analysis (PCA) method. The log-scaled y-axis accentuates the significant reduction in dimensionality from the global searching space $\bm{\Omega}_{GS}$ to the fine-tuning space $\bm{\Omega}_{LF}$, evidencing the constraining influence of the $r_{GS}$ reward function. The relationship between the two training stages is further explored in Fig.~\ref{fig:reward-tsne}(d), where the combined state-action spaces, $\bm{\Omega}_{all}$, including both training stages, are projected onto a two-dimensional space $\bm{\Omega}_{tSNE}$ using the t-SNE method. 
\begin{equation}
\begin{split}
    \bm{\Omega}_{all} = \bm{\Omega}_{GS} \cup \bm{\Omega}_{LF} \cup \bm{\Omega}'_{LF}\\
    t-SNE: \quad \bm{\Omega}_{all} \rightarrow \bm{\Omega}_{tSNE}
\end{split}
\end{equation}
This projection serves to visualize the distribution of test trajectories selected by the DRL agent throughout the training process. It shows that trajectories generated under the same reward function cluster together within this t-SNE transformed space, while those from different reward function stages are markedly separated, except for the initial trajectories which diverge due to an unrefined policy. The contour in Fig.~\ref{fig:reward-tsne}(d) is colored based on the propulsion efficiency $\eta$ at these testing points, revealing a multifaceted landscape of efficiency. This landscape is characterized by multiple local maxima, highlighting the complex nature of the policy learning task. Among these trajectories, those belonging to $\bm{\Omega}_{LF}$, associated with the fine-tuning stage, are proximal to the regions indicative of an optimal policy. In contrast, trajectories from $\bm{\Omega}_{GS}$, representative of the initial global search phase, predominantly occupy regions with lower and more smooth efficiency values. Trajectories from $\bm{\Omega}'_{LF}$, on the other hand, are found mostly in areas marked by high variability in efficiency. In Fig.~\ref{fig:reward-tsne}(e-h), the chronologically colored dots represent the testing trajectories within the t-SNE space $\bm{\Omega}_{tSNE}$, derived from different stages of the training process: $\bm{\Omega}_{GS}$, $\bm{\Omega}_{LF}$, $\bm{\Omega}_{all}$, and $\bm{\Omega}'_{LF}$. The corresponding contours are colored based on the respective reward functions and control goals: $r_{GS}$, $r_{LF}$, and $\eta$. Fig.~\ref{fig:reward-tsne}(e) highlights trajectories from $\bm{\Omega}_{GS}$ gravitating towards areas with high rewards as per $r_{GS}$. An orange dashed box delineates this high-reward zone. Fig.~\ref{fig:reward-tsne}(f) depicts the early phase of the fine-tuning stage, where the trajectories initially follow a path influenced by $r_{GS}$, indicated by an orange arrow. As the training progresses, a gradual shift towards the high-reward areas of $r_{LF}$, closely aligning with regions of high efficiency, becomes evident. Fig.~\ref{fig:reward-tsne}(g) displays all the testing trajectories, offering a comprehensive view of the DRL agent's progression towards the optimal control policy. Fig.~\ref{fig:reward-tsne}(h) reveals a bifurcation in the trajectory paths within $\bm{\Omega}_{LF}\cup\bm{\Omega}'_{LF}$, separated by a notable gap, accentuated by an orange dashed box. It is important to note that the apparent high-reward coloration within this gap is a result of linear interpolation and does not accurately reflect the actual reward landscape. This visual discrepancy is clarified by the consistently high return trajectory shown in Fig.\ref{fig:reward-tsne}(a-b).

\section{Conclusion}
\label{sec:conclusion}
In this work, we introduced and rigorously evaluated a DRL training approach: asynchronous parallel training (APT). This novel strategy is specifically engineered to expedite the DRL training process, particularly in scenarios where interaction with time-intensive environments, such as high-fidelity FSI simulations, is required. Our application of APT to complex fish-fin ray control tasks demonstrates its exceptional efficacy. In the thrust maximization scenario, the DRL agent equipped with APT achieved a remarkable $86.6\%$ increase in thrust generation compared to the baseline method. Further, in the pursuit of maximizing propulsion efficiency, we pioneered the "Global Searching and Local Fine-Tuning" (GSLF) methodology. This approach effectively navigates the challenge of approximating non-additive control goals by employing a series of additive reward functions. The successful implementation of GSLF, in conjunction with APT, results in a control policy that matches the peak efficiency achieved by the baseline method. This outcome not only highlights the practicality of GSLF in complex control scenarios but also its potential in broadening the applicability of DRL in various fields. The merit and effectiveness of the proposed APT method is further discussed by comparing it with conventional DRL training schemes within a chaotic system governed by the KS equation. In conclusion, our study advances DRL training for computationally demanding environments, combining APT and GSLF to efficiently train DRL agents for dynamic control of complex systems. This innovation has broad implications for future DRL applications in science and engineering.

\section*{Code availability}
The source code of asynchronous parallel training (APT) will be openly available on GitHub at \url{https://github.com/jx-wang-s-group/APT-RL} upon publication.

\section*{Acknowledgment}
The authors would like to acknowledge the funds from Office of Naval Research under award numbers N00014-23-1-2071 and National Science Foundation under award numbers OAC-2047127.

\section*{Author contributions:} X.Y.L. and J.X.W. contributed to the ideation and design associated with deep reinforcement learning (DRL) part of the work. X.Y.L. performed the research related to DRL. D.B., Q.X. and X.Z. contributed to the research related to fluid-structure interaction (FSI) simulations. J.X W. and X.Z. supervised the research. X.Y.L., D.B. and J.X W. wrote the manuscript. X Z. contributed to manuscript editing.


\end{document}